\documentclass[11pt]{article}

\usepackage{amsmath,amssymb}
\usepackage{graphicx}
\usepackage{citesort}

\newcommand{\Mpi}{M_\pi^2}
\newcommand{\Mpipm}{M_{\pi^\pm}^2}
\newcommand{\Mpip}{M_{\pi^+}^2}
\newcommand{\Mpin}{M_{\pi^0}^2}
\newcommand{\MKpm}{M_{K^\pm}^2}
\newcommand{\MK}{M_K^2}
\newcommand{\MKp}{M_{K^+}^2}
\newcommand{\MKn}{M_{K^0}^2}
\newcommand{\Me}{M_\eta^2}

\newcommand{\Reals}{\protect{\mathbb{R}}}
\newcommand{\M}{\mathcal{M}}
\newcommand{\Order}[1]{\mathcal{O}#1}
\newcommand{\Lagr}{\mathcal{L}}
\newcommand{\sh}{\!\not\!}
\newcommand{\Sh}{\!\not\!\!}

\newcommand{\bea}{\begin{eqnarray}}
\newcommand{\eea}{\end{eqnarray}}
\newcommand{\beq}{\begin{equation}}
\newcommand{\eeq}{\end{equation}}

\newcommand{\no}{\nonumber}
\newcommand{\vs}{\vspace{-2.5mm}}

\begin{document} 

\begin{center}
\Large{\bf An introduction to chiral perturbation theory}

\vspace{5mm}
\large{Bastian~Kubis}

\vspace{3mm}
\footnotesize{Helmholtz-Institut f\"ur Strahlen- und Kernphysik (Theorie),\\
Universit\"at Bonn, Nussallee 14--16, D-53115 Bonn, Germany}

\vspace{4mm}
\begin{abstract}
A brief introduction to the low-energy effective field theory
of the standard model, chiral perturbation theory, is presented.
\end{abstract}

\end{center}


\section{Introduction}

The phenomenology of the strong interactions at low energies, 
where the coupling constant of Quantum Chromodynamics (QCD) 
becomes large and renders perturbation theory useless, 
remains one of the major challenges of modern particle physics.
Only two rigorous approaches to this part of the standard model
are known:  lattice QCD; and the effective field theory called
chiral perturbation theory.  Both also offer the only ways to provide
a firmer foundation for nuclear physics, rooted in the standard model.

These lectures provide an introduction to chiral perturbation theory,
organised as follows.  In Sect.~\ref{sec:EFT}, we present
a few fundamental ideas on effective field theories in general.
Section~\ref{sec:lowenergyQCD} 
introduces the basic concept and construction of chiral Lagrangians;
a first application 
is the determination of the ratios of light quark masses.
In Sect.~\ref{sec:higherChPT}, chiral perturbation theory is extended
to higher orders, in particular, the relation of the quark mass expansion
of the pion mass to pion-pion scattering is discussed in some detail.
Section~\ref{sec:baryonChPT} extends chiral perturbation theory to include
nucleons; the complications in doing loop calculations are explained,
and the quark mass expansion of the nucleon mass is discussed
in relation to pion-nucleon scattering.
A final outlook summarises some major omissions that could not be covered here.

Several useful pedagogical introductions and review articles on the 
subject have been consulted in the course of the preparation of these lectures, see
in particular~\cite{EckerChPT,ManoharEFT,MeissnerChPT,SchererChPT,GasserSchladming,ColangeloZuoz,BMreview},
many of them being much more comprehensive than the present article,
which are therefore recommended for further reading.


\section{Effective field theories}\label{sec:EFT}

The large number of different areas in physics describe phenomena at very disparate
scales of length, time, energy, or mass.  
It is a rather intuitive idea that, as long as one is only interested in 
a particular parameter range, scales much bigger or much smaller than the
ones one is interested in should not influence the description of the system in question
too strongly.  
Indeed, the two fundamental revolutions in physics in the early 20th century 
did not come about earlier because they involve scales far removed from 
our everyday experience:  for velocities far smaller than the speed of light, 
$v \ll c$, relativity effects can safely be ignored; and for energies and time scales 
much larger than Planck's constant, $E \times t \gg \hslash$, quantum effects
are rarely relevant.  

One striking feature in particle physics is the extremely wide range of observed 
particle masses.  Even ignoring neutrinos, the masses of the fermions comprise
nearly six orders of magnitude, ranging from the electron mass $m_e = 0.511$~MeV
to the top quark mass $m_t \approx 180$~GeV.  
On the other hand, we can very well calculate the properties of certain systems, 
say, the spectrum of the hydrogen atom, without having any precise knowledge of $m_t$ at all.  
In this sense, very heavy particles do not seem to have a significant influence
on the description of the system.

These rather informal ideas are explored more systematically in effective field theories.
Let us, as an example, consider a theory with a set of ``light'' degrees of freedom $l_i$
and heavy fields $H_j$, their respective masses  well separated by a scale $\Lambda$, 
\[ m_{l_i} \ll \Lambda \lesssim M_{H_j} ~. \]
For energies well below $\Lambda$, the heavy particles can be integrated out of the
generating functional, leaving behind an effective Lagrangian for the light degrees of 
freedom only, 
\beq
\Lagr(l_i,H_j) \stackrel{E\ll \Lambda}{\longrightarrow} \Lagr_{\rm eff}(l_i) 
= \Lagr_{d\leq 4} + \sum_{d>4}\frac{1}{\Lambda^{d-4}} \sum_{i_d} g_{i_d}O_{i_d}
~, 
\eeq
i.e.\ the procedure potentially generates non-renormalisable operators of dimension
larger than 4.  
In a so-called ``decoupling'' effective field theory,
the effects of the heavy fields $H_j$ enter $\Lagr_{\rm eff}(l_i)$ either as 
renormalisation effects of the effective coupling constants, 
or as new, higher-dimensional operators suppressed by inverse powers 
of the heavy mass $M_{H_j}$.

We want to briefly illustrate such decoupling effective theories with
a classic example:  
electrodynamics far below the electron mass.


\subsection{An example: light-by-light scattering}

\begin{figure} 
\begin{center}
\includegraphics[width=3.cm]{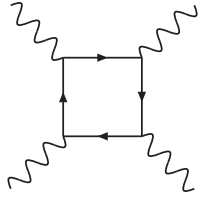}
\caption{Feynman diagram for light-by-light scattering.~\label{fig:lightbylight}}
\end{center} \vspace{-3mm}
\end{figure}

In Quantum Electrodynamics (QED), the only available mass scale is the mass of the electron $m_e$.
If we consider QED at energies far below this mass scale, $\omega \ll m_e$, we should
be able to write down an effective Lagrangian that contains only photons as dynamical degrees
of freedom, 
\[ \Lagr_{\rm QED}[\psi,\bar\psi,A_\mu] \to \Lagr_{\rm eff}[A_\mu] ~.\]
However, the electrons present in the full theory generate effective interactions
for the photon fields, contributing e.g.\ to photon--photon scattering, see 
Fig.~\ref{fig:lightbylight}.
However, we do not have to calculate the underlying loop diagrams explicitly
in order to understand the structure of the effective theory; rather we can
write it down directly in terms of the invariants 
$F_{\mu\nu} F^{\mu\nu} \propto \vec E^2-\vec B^2$,
$F_{\mu\nu} \tilde F^{\mu\nu} \propto \vec E \cdot \vec B$.  
Considering only terms with up to four photon fields, we find
\beq
\Lagr_{\rm eff} = \frac{1}{2}\bigl( \vec E^2-\vec B^2\bigr) + \frac{e^4}{16\pi^2 m_e^4}
\left[ a \bigl( \vec E^2-\vec B^2\bigr)^2 + b \bigl( \vec E \cdot \vec B\bigr)^2 \right] + \ldots~,
\label{eq:EulerHeisenberg}
\eeq
where the prefactor of the interaction term is taken from dimensional analysis,
such that the coupling constants $a$ and $b$ are expected to be of order 1.  
These constants can only be calculated explicitly from the underlying theory, with the result
$7a=b=14/45$~\cite{EulerHeisenberg}.

The following points are to be noted from this brief example, which are typical for the 
construction of (low-energy) effective field theories: \vspace{-1mm}
\begin{enumerate}
\item
We have constructed the interaction terms in the Lagrangian based on the
\emph{symmetries} (gauge invariance, Lorentz invariance)
of the underlying theory, which ought to be shared by the effective one. \vspace{-1mm}
\item
We could only guess the order of magnitude of the effective coupling constants $a$, $b$
correctly, but their exact values are not determined by symmetry considerations alone;
they have to be calculated explicitly from the dynamics of the fundamental theory.\vspace{-1mm}
\item
By considering only the simplest invariant terms that can be constructed in terms of the
field strength tensor $F_{\mu\nu}$ (and its adjoint) and no additional derivatives, 
we have implicitly performed a low-energy expansion of the amplitude, i.e.\ an expansion in powers of
$(\omega/m_e)^{2n}$.\vspace{-1mm}
\item
It is obvious from \eqref{eq:EulerHeisenberg} that the calculation of cross sections
etc.\ is much simpler and more efficiently done using $\Lagr_{\rm eff}$ than performing
the calculation in full QED.
\end{enumerate}


\subsection{Weinberg's conjecture}

The following statement by Weinberg lies at the very heart of the successful
application of effective field theories, using an effective Lagrangian framework:
\begin{center}
\begin{minipage}{9.5cm}
Quantum Field Theory has no content besides unitarity, analyticity, 
cluster decomposition, and symmetries.~\cite{Weinberg79}
\end{minipage}
\end{center}
This means that in order to calculate the S-matrix for any theory
below some scale, simply use the most
general effective Lagrangian consistent with
these principles in terms of the appropriate  
asymptotic states.
This is what we have done in the previous subsection for QED at very low
energies; and we will now follow this principle in the construction of
an effective theory for the strong interactions.


\section{The strong interactions at low energies}\label{sec:lowenergyQCD}
\subsection{The symmetries of Quantum Chromodynamics}

The spectrum of states of strongly interacting particles displays some interesting
features.  Above a typical hadronic mass scale of about 1~GeV, there is a large number
of states, both meson resonances and baryons.  Only a very few (pseudoscalar) states, however, 
are significantly lighter than this mass scale: in particular the pions ($M_\pi \approx 140$~MeV), 
but also kaons ($M_K \approx 495$~MeV) and the eta ($M_\eta \approx 550$~MeV).

The widely accepted theory of the strong interactions is Quantum Chromodynamics,
a theory formulated in terms of quark and gluon fields built on the principle of
colour gauge invariance with the gauge group SU(3)$_c$.  The running strong coupling constant
leads to the phenomena of asymptotic freedom in the high-energy regime, but also to confinement
of the quark and gluon degrees of freedom inside colour-neutral hadronic states.  
The large coupling prevents a use of perturbation theory at low energies, so there is 
no direct and obvious link between QCD and its \emph{fundamental} degrees of freedom, 
and the \emph{relevant} hadronic degrees of freedom as observed in the spectrum of mesons 
and baryons.

\begin{sloppypar}
In order to construct a low-energy effective theory for the strong interactions, 
we have to investigate the symmetries of the QCD Lagrangian more closely.
For this purpose, we decompose the quark fields into its chiral components according to
\beq
q = \frac{1}{2}(1-\gamma_5)q + \frac{1}{2}(1+\gamma_5)q
= P_{L} q + P_{R} q ~=~ q_{L} + q_{R} ~. 
\eeq
Using this, we can write the QCD Lagrangian as
\bea
\Lagr_{\rm QCD} &=& \Lagr_{\rm QCD}^0 - \Lagr_{\rm QCD}^m + \ldots  ~, \no\\
\Lagr_{\rm QCD}^0 &=& -\frac{1}{2} {\rm Tr} \, G_{\mu\nu}^a G^{\mu\nu,a} + 
i \bar q_L \Sh D q_L + i \bar q_R \Sh D q_R ~, \no\\
\Lagr_{\rm QCD}^m &=& \bar q_L \M q_R + \bar q_R \M^\dagger q_L ~,
\eea
where $D_\mu$ is the covariant derivative, 
$G_{\mu\nu}^a$ the gluon field strength tensor, 
$q$ collects the light quark flavours $q^T = (u,d,s)$,
and $\M = {\rm diag}(m_u,\,m_d,\,m_s)$ is the quark mass matrix.
The ellipse denotes the heavier quark flavours, gauge fixing terms etc.
We note that, besides the obvious symmetries like
Lorentz-invariance, ${\rm SU}(3)_c$ gauge invariance, 
and the discrete symmetries $P$, $C$, $T$, $\Lagr_{\rm QCD}$ displays
a \emph{chiral} symmetry in the limit of vanishing quark masses (which is hence
called ``chiral limit''):
$\Lagr_{\rm QCD}^0$ is invariant under chiral $U(3)_L \times U(3)_R$ flavour transformations, 
\beq
(q_L, q_R) \longmapsto (L q_L, R q_R) ~, \quad L,R \in {\rm U}(3)_{L,R} ~.
\eeq
As the masses of the three light quarks are \emph{small} on the typical 
hadronic scale,
$$
m_{u,d,s} \ll 1~{\rm GeV} \approx \Lambda_{hadr} ~,
$$
there is hope that the real world is not too far from the chiral limit, such that
one may invoke a perturbative expansion in the quark masses.  
The effective theory constructed in the following, based on this idea, is therefore
called ``chiral perturbation theory'' (ChPT).
\end{sloppypar}

If we rewrite the symmetry group according to
\beq {\rm U}(3)_L \times {\rm U}(3)_R = 
{\rm SU}(3)_L \times {\rm SU}(3)_R \times {\rm U}(1)_V \times {\rm U}(1)_A ~,
\eeq
where we have introduced vector $V=L+R$ and axial vector $A=L-R$ transformations,
and consider the Noether currents associated with this symmetry group,
it turns out that the different parts of it are realised in very different ways in nature: \vspace{-1mm}
\begin{itemize}
\item The ${\rm U}(1)_V$ current $V^{\mu,0} = \bar q \gamma_\mu q$,
the quark number or baryon number current, is a conserved current in the standard model.\vspace{-1mm}
\item The ${\rm U}(1)_A$ current $A^{\mu,0} = \bar q \gamma_\mu \gamma_5 q$ is
broken by quantum effects, the  ${\rm U}(1)_A$ anomaly, and 
is not a conserved current of the quantum theory.\vspace{-1mm}
\end{itemize}
As far as the chiral symmetry group ${\rm SU}(3)_L \times {\rm SU}(3)_R$ 
and its conserved currents 
\[
\begin{array}{rclrcl}
V^{\mu,a} &=& \bar q \gamma^\mu \dfrac{\lambda^a}{2} q ~,  & 
\quad \partial_\mu V^{\mu,a} &=& 0 ~, \\[2mm]
A^{\mu,a} &=& \bar q \gamma^\mu \gamma_5 \dfrac{\lambda^a}{2} q ~,  &
\quad \partial_\mu A^{\mu,a} &=& 0 ~, 
\end{array} \quad a=1,\ldots, 8
\]
are concerned,
they are certainly broken \emph{explicitly} by the quark masses, but this is 
expected to be a small effect.  Hence the main question is whether chiral symmetry 
is realised in nature in the Wigner--Weyl mode, i.e.\ the symmetry is manifest
in the spectrum in terms of multiplets, or whether it is realised as the Goldstone
mode, i.e.\ the symmetry is hidden or spontaneously broken.  

Can chiral symmetry of the strong interactions be in the Wigner--Weyl mode?
In this case, the conserved axial charges annihilating the vacuum,
\[ Q^a_5 |0\rangle = 0 ~, \quad  Q_5^a = \int d^3 x A^{0,a}(x) ~,\]
would lead to parity doubling in the hadron spectrum.
Phenomenologically, we find (approximate) SU(3)$_V$ multiplets, but
no parity doubling is observed.  Furthermore, unbroken chiral symmetry
would lead to a vanishing difference of the vector--vector and axial--axial
vacuum correlators, $\langle 0 | VV | 0 \rangle - \langle 0 | AA | 0 \rangle = 0$.
This difference can be measured in hadronic tau decays 
$\tau \to \nu_\tau + n\, \pi$, leading to a non-vanishing result~\cite{Aleph}.

If chiral symmetry is, however, realised in the Goldstone mode, the Vafa--Witten 
theorem~\cite{VafaWitten} asserts that the vector subgroup should remain unbroken,
in accordance with the observation of hadronic multiplets, 
so the symmetry breaking pattern would be
\beq
{\rm SU}(3)_L \times {\rm SU}(3)_R \stackrel{{\rm SSB}}{\longrightarrow} 
{\rm SU}(3)_V ~.
\eeq
The axial charges then commute with the Hamiltonian, but do not leave the ground state invariant.
As a consequence, massless excitations, so-called ``Goldstone bosons'' appear,
which are non-interacting for vanishing energy.
In the case at hand, the 8 Goldstone bosons should be pseudoscalars, which
the lightest hadrons in the spectrum indeed are, namely 
$\pi^\pm$, $\pi^0$, $K^\pm$, $K^0$, $\bar K^0$, and $\eta$. 
The task is now to construct a low-energy theory for these Goldstone bosons.
This is an example for a \emph{non-decoupling} effective field theory:
in contrast to the example of QED at energies below the electron mass described earlier,
the transition from the full to the effective theory proceeds via a phase transition / 
via spontaneous symmetry breakdown, in the course of which new light degrees of 
freedom are generated.


\subsection{Construction of the effective Lagrangian}

We want to develop a general formalism~\cite{CCWZ} to construct the effective theory
for the Goldstone bosons corresponding to a symmetry group $G$ spontaneously
broken to its subgroup $H$, 
hence  $G \stackrel{\rm SSB}{\longrightarrow} H$, where
${\rm dim}(G) - {\rm dim}(H) = n$.
We combine the Goldstone boson fields in a vector
$\vec \phi = (\phi_1, \ldots, \phi_n)$, $\phi_i : ~ M^4 \to \Reals$
(where $M^4$ denotes Minkowski space).
The symmetry group $G$ acts on $\vec \phi$ according to
\[ g\in G\, : \quad \vec\phi \longmapsto \vec \phi' = \vec f\bigl(g,\vec \phi\bigr) ~, \]
which has to obey the composition law
\[ \vec f\bigl(g_1, \vec f(g_2,\vec \phi)\bigr) = \vec f\bigl(g_1 g_2, \vec \phi\bigr) ~. \]
Consider the image of the origin $\vec f(g, \vec 0)$:
elements leaving the origin invariant form a subgroup, the conserved subgroup $H$.
Now we have
\[ 
\forall g\in G \quad \forall h\in H  \quad \vec f(gh,\vec 0) = \vec f(g,\vec 0) ~,
\]
therefore $\vec f$ maps the quotient space $G/H$ onto the space of Goldstone boson fields.
This mapping is invertible, as  
$ \vec f(g_1, \vec 0) = \vec f(g_2,\vec 0) $ implies $g_1 g_2^{-1} \in H$.
Hence we conclude that the
Goldstone bosons can be identified with elements of $G/H$.
For $q_i \in G/H$, the action of $G$ on $G/H$ is then given by 
\[ g q_1 = q_2 h(g, \, q_1) ~, \quad h(g,\, q_1) \in H  ~,\]
therefore the coordinates of $G/H$ transform nonlinearly under $G$.

In the case of QCD, we denote the group elements by $g \sim (g_R,\, g_L)$,
$g_{R/L} \in {\rm SU}(3)_{R/L}$,
with the composition law
\[ g_1 g_2 = (g_{R_1},\,g_{L_1})(g_{R_2},\,g_{L_2}) = 
(g_{R_1} g_{R_2}, \, g_{L_1} g_{L_2} ) ~.\]
The choice of a representative element inside each equivalence class is in principle arbitrary, 
the convention for $g H \in G/H$ is to rewrite 
$(g_R,\,g_L) = (\mathbb{E},\, {g_L g_R^{-1}}) (g_R,\,g_R)$ and to characterise 
each element of $G/H$, i.e.\ each Goldstone boson, uniquely by a unitary matrix
\beq 
U = g_L g_R^{-1} = \exp \left( \frac{i\lambda_a {\phi_a}}{{F'}} \right) ~. 
\eeq
Here, the $\phi_a$, $a=1,\ldots,8$ are the Goldstone boson fields, and
$F'$ is a dimensionful constant to be determined later.
How does $U$ transform under the chiral group?  We find
\[
(R,L) (\mathbb{E},U) \cdot H =  (\mathbb{E}, LUR^\dagger)(R,R)\cdot H 
= (\mathbb{E}, {LUR^\dagger}) \cdot H ~,
\]
therefore
\beq
U \stackrel{G}{\longmapsto} LUR^\dagger ~.
\eeq


\subsection{The leading-order Lagrangian}
 
Now we know how the Goldstone boson fields transform under chiral transformations,
we can proceed to construct a Lagrangian in terms of the matrix $U$ that is
invariant under ${\rm SU}(3)_L \times {\rm SU}(3)_R$.  
As we want to construct a \emph{low-energy} effective theory,
the guiding principle is to use the power of momenta or derivatives to order 
the importance of various possible terms.  ``Low energies'' here refer to 
a scale well below 1~GeV, i.e.\ an energy region where the Goldstone bosons
are the only relevant degrees of freedom.

Lorentz invariance dictates that Lagrangian terms can only come in even powers 
of derivatives, hence $\Lagr$ is of the form
\beq \Lagr = \Lagr^{(0)} + \Lagr^{(2)} + \Lagr^{(4)} + \ldots ~.\eeq
However, as $U$ is unitary, therefore $U U^\dagger = \mathbb{E}$, 
$\Lagr^{(0)}$ can only be a constant.  
Therefore, in accordance with the Goldstone theorem, the leading term in the Lagrangian
is $\Lagr^{(2)}$, which already involves derivatives.  It can be shown to consist of 
one single term,
\beq 
\Lagr^{(2)} = \frac{F^2}{4} \langle \partial_\mu U \partial^\mu U^\dagger \rangle ~, \label{eq:L2}
\eeq
where $F$ is another dimensionful constant, and 
\beq
U=\exp \left(\frac{i\phi}{F'}\right) ~, \quad
\phi = \sqrt{2} \left( \begin{array}{ccc} 
\frac{\phi_3}{\sqrt{2}} + \frac{\phi_8}{\sqrt{6}} & \pi^+& K^+ \\
\pi^- & -\frac{\phi_3}{\sqrt{2}} + \frac{\phi_8}{\sqrt{6}} & K^0 \\
K^- & \bar K^0 & -\frac{2\phi_8}{\sqrt{6}} \end{array} \right) ~.
\eeq
Expanding $U$ in powers of $\phi$, $U=1+i\phi/F'-\phi^2/(2F'^2)+\ldots\,$,
we find the canonical kinetic terms
\[ \Lagr^{(2)} = \partial_\mu \pi^+ \partial^\mu \pi^- + \partial_\mu K^+ \partial^\mu K^- + \ldots \]
exactly for $F'=F$.
The invariance of \eqref{eq:L2} under ${\rm SU}(3)_L \times {\rm SU}(3)_R$ is
easily verified:
\beq
\langle \partial_\mu U \partial^\mu U^\dagger \rangle ~\longmapsto~
\langle \partial_\mu U' \partial^\mu U'^\dagger \rangle 
~=~ \langle L \partial_\mu U R^\dagger R \partial^\mu U^\dagger L^\dagger \rangle
~=~  \langle \partial_\mu U \partial^\mu U^\dagger \rangle ~.\no
\eeq

We remark here that our derivation of \eqref{eq:L2} is somewhat heuristic.
A more formal proof of the equivalence of QCD and its representation in terms
of an effective Lagrangian, based on an analysis of the chiral Ward identities,
is given in~\cite{LeutFound}.  Furthermore, we have neglected anomalies in the 
above reasoning, which can be shown to enter only at next-to-leading order~\cite{WZW}.


\subsection{The constant \boldmath{$F$}}

In order to determine the constant $F$, we proceed 
to calculate the Noether currents $V^\mu_a$, $A^\mu_a$ from $\Lagr^{(2)}$:
\beq
V^\mu_a, A^\mu_a  ~=~ R^\mu_a \pm L^\mu_a ~=~ i\, \frac{F^2}{4} 
\langle \lambda_a \bigl[ \partial^\mu U, U^\dagger\bigr]_\mp \rangle  ~.
\eeq
Expanding the axial current in powers of $\phi$, we find 
$A^\mu_a = -F \partial^\mu \phi_a + \Order(\phi^3)$, such that
we can calculate the matrix element of the axial current between a one-boson state
and the vacuum, 
\beq \langle 0 | A^\mu_a | \phi_b(p) \rangle = i p^\mu \delta_{ab} F  ~,
\eeq
from which we conclude that $F$ is the pion (meson) decay constant 
(in the chiral limit),
which is measured in pion decay $\pi^+ \to \ell^+ \nu_\ell\,$,
$
F \approx F_\pi = 92.4~{\rm MeV} \,.
$


\subsection{Explicit symmetry breaking:  quark masses}

So far, we have only considered the chiral limit $m_u=m_d=m_s=0$. 
Accordingly, we have constructed a 
theory for {massless} Goldstone bosons, and indeed, 
$\Lagr^{(2)}$ does not contain any mass terms.
In nature, the quark masses are small, but certainly non-zero,
therefore chiral symmetry is explicitly broken.
In order to account for this fact, the quark masses 
have to be re-introduced perturbatively.
For this purpose, we have to understand
the transformation properties of the symmetry breaking term;
then the (appropriately generalised) effective Lagrangian is still the right tool
to systematically derive all symmetry relations of the theory.

From the QCD mass term
$
\Lagr_{\rm QCD}^m = \bar q_L \M q_R + \bar q_R \M^\dagger q_L \,,
$
we notice it \emph{would} be invariant under chiral transformations
if $\M$ transformed according to
\beq
\M \longmapsto \M' = L \M R^\dagger  ~.
\eeq
Assuming this, we now construct chirally invariant Lagrangian terms
from $U$, derivatives thereon, plus the quark mass matrix $\M$;
this procedure guarantees that chiral symmetry is broken in
exactly the same way in the effective theory as it is in QCD.

\begin{sloppypar}
At leading order, i.e.\ to linear order in the quark masses and
without any further derivatives, we find exactly one term in the chiral Lagrangian,
such that $\Lagr^{(2)}$ is of the form
\beq
\Lagr^{(2)} = \frac{F^2}{4} \langle \partial_\mu U \partial^\mu U^\dagger 
+2B \bigl( \M U^\dagger + \M^\dagger U \bigr) \rangle ~. \label{eq:L2mass}
\eeq
Expanding once more in powers of $\phi$, we can read off the mass terms and find
\beq
\Mpipm = B(m_u+m_d) ~, \quad
\MKpm  = B(m_u+m_s) ~, \quad
\MKn   = B(m_d+m_s) ~.
\eeq
We recover the Gell-Mann--Oakes--Renner relation 
$ M_{GB}^2 \propto m_q$, which justifies the
unified power counting for the expansion in numbers of derivatives as well as quark masses
according to $m_q = \Order(p^2)$.
We furthermore find that the 
flavour-neutral states $\phi_3$, $\phi_8$ are mixed when isospin breaking due
to a difference in the light quark masses is allowed for, $m_u-m_d\neq 0$:
\[ \Lagr^{(2)} ~\longrightarrow~ 
\frac{B}{2} \left(\!\! \begin{array}{c} \phi_3 \\ \phi_8 \end{array} \!\!\right)^T
\left(\! \begin{array}{cc}
m_u+m_d & \frac{1}{\sqrt{3}}(m_u-m_d) \\
\frac{1}{\sqrt{3}}(m_u-m_d) & \frac{1}{3}(m_u+m_d+4m_s)
\end{array} \!\right)
\left(\!\! \begin{array}{c} \phi_3 \\ \phi_8 \end{array} \!\!\right) ~,
\]
which can be diagonalised by the rotation
\[
\left(\!\! \begin{array}{c} \pi^0 \\ \eta \end{array} \!\!\right) = 
\left(\! \begin{array}{cc}
\cos\epsilon & -\sin\epsilon \\
\sin\epsilon & \cos\epsilon
\end{array} \!\right)
\left(\!\! \begin{array}{c} \phi_3 \\ \phi_8 \end{array} \!\!\right) ~, \quad
\epsilon = \frac{1}{2} \arctan \biggl( 
\frac{\sqrt{3}}{2} \frac{{m_d-m_u}}{m_s-\hat m} \biggr) ~,  
\]
where $\hat m = (m_u+m_d)/2$.
The mass eigenvalues receive corrections to the isospin limit,
which are however of second order in $m_u-m_d$, 
\bea
\Mpin &=& B({m_u}+{m_d}) - \Order\bigl((m_u-m_d)^2\bigr) ~,\no\\
\Me &=& \dfrac{B}{3}({m_u}+{m_d}+4{m_s}) + \Order\bigl((m_u-m_d)^2\bigr) ~.
\eea
In the isospin limit, we of course find $\Mpipm = \Mpin$, $\MKpm = \MKn$. 
Finally, we can deduce the Gell-Mann--Okubo mass formula (for the pseudoscalars)
\beq 4\MK = 3\Me+\Mpi ~,\eeq
which is found to be fulfilled in nature to 7\% accuracy.
\end{sloppypar}


\subsection{Quark mass ratios}

The unknown factor $B$ in the Gell-Man--Oakes--Renner relations
prevents a direct quark mass determination from pseudoscalar meson masses.
However, we can form quark mass ratios in which $B$ cancels:
\bea
{\frac{m_u}{m_d}} &=& \frac{\MKp-\MKn+\Mpip}{\MKn-\MKp+\Mpip} ~\approx~{0.66} ~, \label{eq:mumd}\\
{\frac{m_s}{m_d}} &=& \frac{\MKn+\MKp-\Mpip}{\MKn-\MKp+\Mpip} ~\approx~ {22} ~.\label{eq:msmd}
\eea
In particular the result for
$m_u/m_d$ is remarkable:  it is very different from 1, so
why is there no large isospin violation observed in nature?
The answer is threefold:
first, in purely pionic physics, only $(m_d-m_u)^{{2}}$ occurs, 
hence strong isospin violation is of second order.
Second, $(m_d-m_u)/m_s~$ (as showing up e.g.\ in the $\pi^0\eta$ mixing angle $\epsilon$) is small;
and third, compared to the typical hadronic scale, $(m_d-m_u)/\Lambda_{\rm hadr}~$ is small, too.

We can calculate the (strong) pion mass difference 
\[ 
\Mpin = \Mpip \left\{ 1-\frac{(m_d-m_u)^{{2}}}{8\hat m(m_s-\hat m)} + \ldots \right\} ~,
\]
and evaluate it numerically by plugging in the quark mass ratios to find
\beq 
M_{\pi^+} - M_{\pi^0} \approx {0.1~{\rm MeV}} ~, 
\eeq
while the experimental mass difference is
$\left(M_{\pi^+} - M_{\pi^0}\right)_{\rm exp} \approx 4.6~{\rm MeV} $.
The difference, the by far larger effect, is due to the second source
of isospin violation that we have neglected so far: electromagnetism.

\subsection{Electromagnetic effects}

The coupling of $\Lagr^{(2)}$ to external vector ($v_\mu$) and axial vector ($a_\mu$) currents 
is rather straightforward: we only have to replace the ordinary derivative
by a covariant one according to
\beq 
\partial_\mu U \to D_\mu U = \partial_\mu U - i [v_\mu,U] - i \{ a_\mu,U\} ~.
\eeq
If we insert the photon field for the vector current, $v_\mu = e A_\mu$, this
will generate all the couplings necessary to calculate, say, the electromagnetic
form factor of the pion, or pion Compton scattering.

However, including electromagnetism via minimal substitution alone does not
generate the most general effects due to \emph{virtual} photons.
Consider, e.g., the contribution of a photon loop to the pion self-energy diagram,
Fig.~\ref{fig:photon}:
\begin{figure}
\begin{center}
\includegraphics[height=1.5cm]{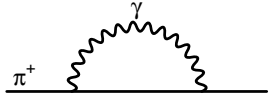}
\caption{Photon loop diagram contributing to the pion self energy.\label{fig:photon}}
\end{center} \vspace{-3mm}
\end{figure}
for dimensional reasons, the contribution to the pion mass has to vanish in the chiral limit,
while a non-vanishing term can be generated in certain models.  
Naively speaking, Fig.~\ref{fig:photon} neglects photon exchanges 
between the (charged) quarks \emph{inside} the pion.
We therefore have to generalise the chiral Lagrangian once more.
We proceed~\cite{Urech} in analogy to the quark mass term, and 
now include the quark charge matrix as an additional element, 
$Q = e\; {\rm diag}\left(2,-1,-1\right)/3$.
The part of the QCD Lagrangian coupling quarks to photons, 
decomposed into chiral components, takes the form
\beq 
\Lagr_{\rm QCD}^{\rm em} = - \bar q Q \Sh A q ~\longrightarrow~
- \bar q_L Q_L \Sh A q_L - \bar q_R Q_R \Sh A q_R ~. \label{eq:LQCDem}
\eeq 
If we postulate the following transformation law(s) for the
spurion fields $Q_{L,R}$:
\[ Q_L \longmapsto L Q_L L^\dagger ~, \quad Q_R \longmapsto R Q_R R^\dagger ~,\]
then \eqref{eq:LQCDem} is seen to be invariant under chiral transformations.
We hence construct Lagrangian terms using $Q_{L,R}$, and 
set $Q_L = Q_R = Q$ in the end.
The power counting is generalised to count  $Q_{L,R} = \Order (p)$.
We find one single term at $\Order(e^2) = \Order(p^2)$: 
\beq \Lagr_{\rm em}^{(2)} ~=~ C \langle Q_L U Q_R U^\dagger \rangle \label{eq:L2em} ~,\eeq
which contributes to the masses of the charged mesons:
\beq
\left(\Mpip-\Mpin\right)_{\rm em} = \left(\MKp-\MKn\right)_{\rm em} 
= \frac{2Ce^2}{F^2} ~. \label{eq:Dashen}
\eeq
The equality of electromagnetic contributions to pion and kaon mass differences in the chiral limit
is known as Dashen's theorem~\cite{Dashen}.
\eqref{eq:L2em} has no contributions to neutral masses, or to $\pi^0\eta$-mixing.
With electromagnetic effects included, we find an improved quark mass ratio
\beq
\frac{m_u}{m_d} ~=~ \frac{\MKp-\MKn+2\Mpin-\Mpip}{\MKn-\MKp+\Mpip} ~=~ 0.55 ~,
\eeq
which deviates significantly from \eqref{eq:mumd}.


\subsection{\boldmath{$\pi\pi$} scattering to leading order}

With the constants $F$, $B$ (in products with quark masses), $C$ fixed from phenomenology, 
the leading-order Lagrangian~\eqref{eq:L2mass}, \eqref{eq:L2em} is completely determined, and we can go on and 
make predictions for other processes.  A particularly important example is
pion-pion scattering.  
For now, we revert to the isospin limit and set $m_u=m_d$, $e^2=0$,
such that the scattering amplitude can be decomposed as
\beq
M(\pi^a \pi^b \to \pi^c \pi^d ) = \delta^{ab}\delta^{cd} A(s,t,u) 
+  \delta^{ac}\delta^{bd} A(t,u,s) 
+  \delta^{ad}\delta^{bc} A(u,s,t) ~.
\eeq
If we calculate the invariant amplitude $A(s,t,u)$ from $\Lagr^{(2)}$, we find
\beq 
A(s,t,u) = \frac{s-\Mpi}{F^2} ~, \label{eq:pipiCA}
\eeq
a parameter-free prediction~\cite{Weinberg66}.
The isospin amplitudes are then given by
\bea
T^{I=0} &=& 3A(s,t,u)+A(t,u,s)+A(u,s,t) ~,\no\\
T^{I=1} &=& A(t,u,s)-A(u,s,t) ~,\no\\
T^{I=2} &=& A(t,u,s)+A(u,s,t) ~.
\eea
If we furthermore define the $s$-wave scattering lengths, proportional 
to the scattering amplitudes at threshold, 
$ a_0^I = T^I (s=4\Mpi,t=u=0)/32\pi $,
we find
\beq
a_0^0 = \frac{7\Mpi}{32\pi F_\pi^2} = 0.16 ~, \quad
a_0^2 = - \frac{\Mpi}{16\pi F_\pi^2} = -0.045 ~. \label{eq:a0Op2}
\eeq


\section{Chiral perturbation theory at higher orders}\label{sec:higherChPT}

So far, we have only considered chiral Lagrangians at leading order, i.e.\ $\Order(p^2)$.
Are there good reasons to go beyond that level?
First of all, although $\Order(p^0)$ interactions are forbidden by chiral symmetry, 
all higher orders are allowed and therefore present in principle.  
They ought to be smaller at low energies, but for precision predictions,
these corrections should be taken into account.
Second, although ChPT is an effective field theory, it is still a quantum theory,
i.e.\ we should also expect loop contributions.
Remembering the $\pi\pi$ scattering amplitude \eqref{eq:pipiCA}, we notice
that it is \emph{real}, while unitarity requires the partial waves $t_\ell^I$
to obey
\beq 
{\rm Im} \,t_\ell^I = \sqrt{1-\frac{4\Mpi}{s}} \left| t_\ell^I \right|^2 ~.
\eeq
The correct imaginary parts are only generated perturbatively by loops.
But how do loop diagrams feature in the power counting scheme?
What about divergences arising thereof, how does renormalisation work in such a theory?


\subsection{Weinberg's power counting argument}

\begin{figure}
\begin{center}
\includegraphics[width=0.75\linewidth]{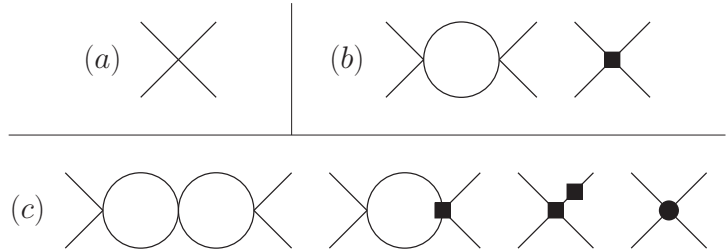}
\caption{Feynman graphs contributing to $\pi\pi$ scattering
at (a)~$\Order(p^2)$, (b)~$\Order(p^4)$, (c)~$\Order(p^6)$.
The squares denote vertices from $\Lagr^{(4)}$, the circle
a vertex from $\Lagr^{(6)}$.
\label{fig:pipiOp246}}
\end{center} \vspace{-3mm}
\end{figure}

Let us consider an arbitrary loop diagram based on the general effective Lagrangian
$ \Lagr_{\rm eff} = \sum_d \Lagr^{(d)}$, where $d$ denotes the chiral power of the
various terms.
If we calculate a diagram
with $L$ loops, $I$ internal lines, and $V_d$ vertices of order $d$, 
the generic form of the corresponding amplitude ${\cal A}$ 
in terms of powers of momenta is
\beq 
{\cal A} \propto \int (d^4p)^L \frac{1}{(p^2)^I} \prod_d (p^d)^{V_d} ~.
\eeq
Let ${\cal A}$ be of chiral dimension $\nu$, then obviously
$ \nu = 4L-2I + \sum_d d V_d \,.
$
We use the topological identity 
$ L = I - \sum_d V_d + 1 $
to eliminate $I$ and find
\beq
\nu = \sum_d V_d(d-2) + 2L+2 ~. \label{eq:powercounting}
\eeq
The following points are to be noted about \eqref{eq:powercounting}: \vspace{-1.mm}
\begin{itemize}
\item The chiral Lagrangian starts with $\Lagr^{(2)}$, i.e.\ $d \geq 2$, 
therefore the right-hand-side of \eqref{eq:powercounting} is a sum of non-negative terms.
Consequently, for fixed $\nu$, there is only a finite number of combinations $L$, $V_d$
that can contribute. \vspace{-1.mm}
\item Each additional loop integration suppresses the amplitude by two orders
in the momentum expansion. \vspace{-1.mm}
\end{itemize}
As an example, let us consider $\pi\pi$ scattering.
At lowest order $p^2$, only tree-level graphs composed 
of vertices of $\Lagr^{(2)}$ contribute ($V_{d>2}=0$, $L=0$).
The only graph is shown in Fig.~\ref{fig:pipiOp246}(a).
At $\Order(p^4)$, there are two possibilities:
either one-loop graphs composed only of lowest-order vertices ($V_{d>2}=0$, $L=1$),
or tree graphs with exactly one insertion from $\Lagr^{(4)}$ ($V_4=1$, $V_{d>4}=0$, $L=0$).
Example graphs for both types are given in Fig.~\ref{fig:pipiOp246}(b).
Finally, at $\Order(p^6)$, \eqref{eq:powercounting} allows for four different types
of graphs: two-loop graphs with $\Lagr^{(2)}$ vertices ($V_{d>2}=0$, $L=2$);
one-loop graphs with one vertex from  $\Lagr^{(4)}$ ($V_4=1$, $V_{d>4}=0$, $L=1$);
tree graphs with two insertions from  $\Lagr^{(4)}$ ($V_4=2$, $V_{d>4}=0$, $L=0$);
and tree graphs with one insertion from  $\Lagr^{(6)}$ ($V_4=0$, $V_6=1$, $V_{d>6}=0$, $L=0$).
Typical examples for all these are displayed in Fig.~\ref{fig:pipiOp246}(c).


\subsection{Chiral symmetry breaking scale \boldmath{$\Lambda_\chi$}}

Although we have now established a power counting scheme that determines
the power of momenta $p^n$ at which a certain diagram contributes, it is not
yet clear compared to what \emph{scale} these momenta are to be small.
If we write the effective Lagrangian in the (slightly unconventional) form
\beq
\Lagr_{\rm eff} = \dfrac{F^2}{4} \left\{ \langle \partial_\mu U \partial^\mu U^\dagger \rangle
+ \dfrac{1}{{\Lambda_\chi^2}} \tilde \Lagr^{(4)} + \dfrac{1}{{\Lambda_\chi^4}} \tilde \Lagr^{(6)}
+\ldots \right\} ~,
\eeq
we need to know what  the chiral symmetry breaking scale $\Lambda_\chi$ is.
In other words, if we calculate higher-order corrections, what is their expected size 
$\propto p^2/\Lambda_\chi^2$, $\propto \Mpi/\Lambda_\chi^2$?

As a first argument, let us compare the two $\Order(p^4)$ contributions
to $\pi\pi$ scattering in Fig.~\ref{fig:pipiOp246}(b) in a generic manner:
\begin{center}
\begin{tabular}{rl}
\begin{minipage}{2cm}
\flushright{\includegraphics[height=8mm]{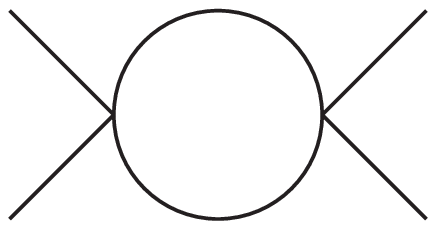}}
\end{minipage}&
$ \propto~ \displaystyle{\int} \dfrac{d^4 p}{(2\pi)^4} \dfrac{1}{(p^2)^2} \biggl(\dfrac{p^2}{F^2}\biggr)^2
\stackrel{\rm dim.reg}{\propto}  \dfrac{p^4}{(4\pi)^2F^4} \log \mu  $ ~,\\[5mm]
\begin{minipage}{2cm}
\flushright{\includegraphics[height=8mm]{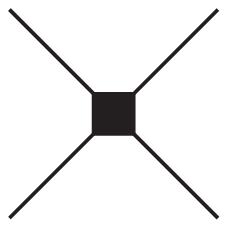} }
\end{minipage}&
$\propto ~ \dfrac{F^2}{\Lambda_\chi^2}\dfrac{p^4}{F^4} \ell_i(\mu)  ~.$
\end{tabular}
\end{center}
The scale-dependent ``low-energy constant'' (LEC)
$\ell_i(\mu)$ multiplying the tree-level graph
compensates for the logarithmic scale dependence of the loop graph
(as evaluated in dimensional regularisation).  
Naturally, the finite part of $\ell_i(\mu)$ should not be expected to be smaller than
the shift induced by a change in the scale $\mu$, therefore 
\beq
\Lambda_\chi \approx 4\pi F \approx 1.2~{\rm GeV} ~.
\eeq
As a second argument, we have to remember that we
have constructed an effective theory for Goldstone bosons, 
which are the only dynamical degrees of freedom.
The effective theory must fail 
once the energy reaches the resonance region, 
hence for
$
p^2/\Lambda_\chi^2 \approx p^2/M_{\rm res}^2 \approx 1
$.
The resonance masses are channel-dependent, the lightest being $M_{\rm res} = M_\rho = 770$~MeV, 
and typically $M_{\rm res} \approx 1$~GeV, therefore this second estimate is
roughly consistent with $\Lambda_\chi = 4\pi F$.


\subsection{The chiral Lagrangian at higher orders}

The number of independent terms and corresponding low-energy constants
increases rapidly at higher orders.  In fact, for chiral SU($n$), 
$n=(2,3)$,\footnote{In contrast to $\Lagr^{(2)}$, which has the same form for both SU(2) and SU(3),
the number of terms at higher orders is different in both theories because,
although both have the same most general SU($N$) Lagrangian, certain
matrix-trace (Cayley-Hamilton) relations render some of the structures redundant,
such that the minimal numbers of independent terms differ.}
\begin{center}
\begin{tabular}{llcll}
$ \Lagr^{(2)}$ &contains& (2,\,2) &constants &($F$, $B$), \\
$ \Lagr^{(4)}$ &contains& (7,\,10) &constants &\cite{GL84,GL85},\\
$ \Lagr^{(6)}$ &contains& (53,\,90) &constants &\cite{BCE99}
\end{tabular}
\end{center}
(discounting so-called contact terms that depend on external fields only).
As an example, we display $\Lagr^{(4)}$ explicitly for chiral SU(3):
\bea
\Lagr^{(4)} \!\!&=&\!\! {L_1} \langle D_\mu U^\dagger D^\mu U\rangle^2
+ {L_2} \langle D_\mu U^\dagger D_\nu U \rangle \langle D^\mu U^\dagger D^\nu U \rangle \no\\
&+&\!\! {L_3} \langle D_\mu U^\dagger D^\mu U D_\nu U^\dagger D^\nu U\rangle 
+{L_4} \langle D_\mu U^\dagger D^\mu U\rangle\langle \chi^\dagger U + \chi U^\dagger \rangle \no\\
&+&\!\! {L_5} \langle D_\mu U^\dagger D^\mu U (\chi^\dagger U + \chi U^\dagger) \rangle 
+ {L_6} \langle \chi^\dagger U + \chi U^\dagger \rangle^2 \no\\
&+&\!\! {L_7} \langle \chi^\dagger U - \chi U^\dagger \rangle^2 
+ {L_8}  \langle \chi^\dagger U\chi^\dagger U + \chi U^\dagger\chi U^\dagger\rangle \no\\
&-&\!\! i\,{L_9} \langle F_R^{\mu\nu} D_\mu U D_\nu U^\dagger
+  F_L^{\mu\nu} D_\mu U^\dagger D_\nu U\rangle
+{L_{10}} \langle U^\dagger F_R^{\mu\nu} U F_{L \mu\nu} \rangle ~.~ \label{eq:L4SU3}
\eea
Here, $\chi = 2B(s+ip)$ collects (pseudo)scalar source terms, where $s$ contains
the quark mass matrix, $s=\M+\ldots\,$;
vector and axial currents are combined as $r_\mu = v_\mu+a_\mu$, $l_\mu = v_\mu-a_\mu$,
from which one can form field strength tensors,
$F_R^{\mu\nu} = \partial^\mu r^\nu -\partial^\nu r^\mu -i[r^\mu,r^\nu]$, 
and similarly $F_L^{\mu\nu}$. 
We note that $L_{1-3}$ multiply structures containing four derivatives; 
$L_{4,5}$ those with two derivatives and one quark mass term;
the structures corresponding to $L_{6-8}$ scale with the quark masses squared.
$L_{9,10}$ only contribute to observables with external vector and axial vector sources.
The seven similar constants in SU(2) that we will also partly use later
are conventionally denoted by ${\ell_i}$, $i=1,\ldots ,7$.

\subsection{The physics behind the low-energy constants}

To better understand the role of the low-energy constants and the physics they incorporate,
let us consider massive states (resonances) that are ``integrated out''
of the theory, i.e.\ no dynamical degrees of freedom for energies
below $\Lambda_\chi$:
\beq 
\Lagr[U,\partial U,\ldots,{H}] ~\stackrel{\Lambda}{\longrightarrow} ~
\Lagr_{\rm eff} [ U,\partial U,\ldots] ~.
\eeq
A Lagrangian for the resonance fields, coupled to source terms, 
is of the form
\beq 
\Lagr[H] = \frac{1}{2} \left( \partial_\mu H \partial^\mu H - M_H^2 H^2 \right) + J H ~,
\eeq
where $J$ is a  current formed of light degrees of freedom.
In the path integral, we can use the fact
the the heavy particle propagator $\Delta_H(x-y)$
is peaked for small separations $|x-y|$, 
\beq
S_{\rm eff}[J] = 
-\frac{1}{2} \int d^4x d^4y \,J(x) \Delta_H(x-y)  J(y) 
\simeq \int d^4 x \frac{1}{2{M_H^2}} J(x) J(x) + \ldots ~,
\eeq
i.e.\ this generates local higher-order operators, with the 
couplings proportional to the inverse heavy masses.
We conclude therefore that the effects of higher-mass states
on the Goldstone boson interactions are hidden in the low-energy constants of higher-order
terms in the chiral Lagrangian.

As an example, we consider the pion vector form factor $F_\pi^V(s)$, defined as
\beq 
\langle \pi^a(p)\pi^b(p') \,|\, \bar q \frac{\tau^3}{2} \gamma_\mu q \,|\, 0 \rangle
= i \epsilon^{a3b}(p'-p)_\mu F_\pi^V(s) ~, \quad s=(p+p')^2 ~.
\eeq
In ChPT at $\Order(p^4)$, there are loop diagrams contributing to $F_\pi^V(s)$
as well as a tree graph proportional to the low-energy constant $\bar \ell_6$.
Expanding $F_\pi^V(s)$ for small $s$, we find for the radius term 
$\langle r^2\rangle_\pi^V$
\beq 
F_\pi^V(s) = 1 + \frac{1}{6}{\langle r^2\rangle_\pi^V} s + \Order(s^2) ~, \quad
\langle r^2\rangle_\pi^V = \frac{1}{(4\pi F_\pi)^2} ({\bar \ell_6} - 1) ~.
\eeq
Now consider the contribution of the $\rho$-resonance 
(see e.g.~\cite{EGLPdR} for how to treat vector mesons
in the context of chiral Lagrangians) to this form factor, 
\begin{figure}
\begin{center}
\includegraphics[width=7cm]{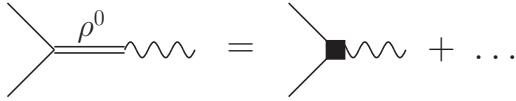}
\caption{The contribution of the $\rho$-resonance to the pion vector form factor
can, at small $t$, be represented by a point-like counterterm.\label{fig:rhoVff}}
\end{center} \vspace{-3mm}
\end{figure}
as shown in Fig.~\ref{fig:rhoVff}.
Expanding the $\rho$ propagator for $s \ll M_\rho^2$, 
\beq
\frac{s}{M_\rho^2-s} = \frac{s}{M_\rho^2}\biggl(1+\frac{s}{M_\rho^2} + \ldots \biggr) ~,
\eeq
we find that identifying the leading term with the $\bar \ell_6$ contribution
reproduces the empirical value for $\bar \ell_6$ nicely.
This is a modern version of the time-honoured concept of ``vector meson dominance'':
where allowed by quantum numbers, the numerical values of LECs are dominated by the contributions
of vector resonances~\cite{EGPdR,Donoghue}.


\subsection{The pion mass to \boldmath{$\Order(p^4)$}}

\begin{figure}
\begin{center}
\includegraphics[width=6cm]{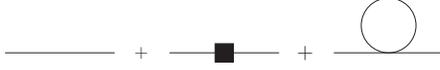} 
\caption{Diagrams contributing to the pion self energy
up to $\Order(p^4)$.\label{fig:pionmass}}
\end{center} \vspace{-3mm}
\end{figure}

As an example for a higher-order calculation, we consider the pion mass up to
$\Order(p^4)$.  The necessary diagrams are shown in Fig.~\ref{fig:pionmass}.
We find for the pion propagator 
\bea
\delta^{ab} \Delta(p) &=& i \int d^4x e^{ipx}
\langle 0 | T \phi^a(x) \phi^b(0) | 0 \rangle  ~, \no\\
\Delta(p) &=& \dfrac{ Z}{\Mpi-p^2} \;(+ {\rm 2~loops}) ~.
\eea
The physical pion mass is given, to this order, by
\beq 
\Mpi = M^2 + \frac{M^2}{2F^2} I + \frac{2{\ell_3}}{F^2} M^4 ~, 
\eeq
where $M^2 = B(m_u+m_d)$, and the loop integral 
\beq  I ~=~\frac{1}{i}\int \frac{d^4 l}{(2\pi)^4} \frac{1}{M^2-l^2} 
\eeq
is actually divergent and has to be regularised.
A ``good'' regularisation scheme, for our purposes, is dimensional regularisation, 
as it preserves all symmetries (which is much more difficult to achieve using a cutoff, say).
In $d$ dimensions, we find
\beq  I ~\to~\frac{1}{i}\int \frac{d^d l}{(2\pi)^d} \frac{1}{M^2-l^2} 
~=~ \frac{M^{d-2}}{(4\pi)^{d/2}} \Gamma\Bigl(1-\frac{d}{2}\Bigr) ~,
\eeq
which is finite for $d\neq 2,\,4,\,6,\ldots\,$, but still divergent for $d=4$:
$$  I ~\to~ \frac{M^2}{8\pi^2} \left\{ {\frac{1}{d-4}} + \log\frac{M}{\mu} + \ldots \right\} ~.
$$
We now tune $\ell_3$ such as to absorb this divergence (as well as the $\mu$-dependence): 
\beq 
\ell_3 ~\to~ - \frac{1}{32\pi^2} \left({\frac{1}{d-4}} + \log\frac{M}{\mu} +\frac{\bar\ell_3}{2}\right)~,
\eeq
such that $\bar \ell_3$ contains the finite part of $\ell_3$, and find
\beq
\Mpi ~=~ M^2 - \dfrac{M^4}{32\pi^2 F^2} {\bar \ell_3} + \Order(M^6) ~. \label{eq:Mpip4}
\eeq

A few comments on renormalisation as we just saw it at work for the first time 
are in order.
The required counterterm to cancel the divergence stems from $\Lagr^{(4)}$;
it is not sufficient to tune $\Lagr^{(2)}$ parameters.
This is typical for a non-renormalisable theory:
going to to higher and higher orders,
we need more and more counterterms.
However, the fact that the theory is non-renormalisable does not mean it is non-calculable,
the only disadvantage is the increasing number of LECs when calculating higher-order corrections.
It is important to note furthermore that the LECs feature in different observables,
and that their divergent parts and scale dependences are always the same.
The renormalisation can be performed on the level of the generating functional
in a manifestly chirally invariant way, and
the $\beta$-functions of the $\ell_i$~\cite{GL84}, $L_i$~\cite{GL85} 
(and also of $\Lagr^{(6)}$ LECs~\cite{BCE00}) are known.
The cancellation of divergences and scale dependence therefore serves as a
powerful check on any specific calculations.


\subsection{Quark mass expansion of the pion mass revisited}

The expression \eqref{eq:Mpip4} provides a
correction to the Gell-Mann--Oakes--Renner relation: generically, it is of the form
\beq 
\Mpi = B(m_u+m_d) + A(m_u+m_d)^2 + \Order(m_q^3)  ~.
\eeq
Apart from naive order-of-magnitude expectations, 
how do we actually know that the leading term dominates?
What if {$\bar \ell_3$} turns out to be anomalously large?
The consequences of this possibility
were explored under the label of ``generalised ChPT'', 
which employs a different power counting scheme~\cite{KMSF95}.
The essential question in order to determine the size of 
second-order corrections in the quark mass expansion of the pion mass is therefore:
how can we learn something about $\bar\ell_3$?


\subsection{\boldmath{$\pi\pi$} scattering at next-to-leading order}

It turns out the process to study is, once more, $\pi\pi$ scattering.
The $\pi\pi$ scattering lengths are known to next-to-next-to-leading order~\cite{BCEGS},
the $\Order(p^4)$ corrections for the $I=0$ scattering length are of the form~\cite{GL84}
\bea
a_0^0 &=& \frac{7\Mpi}{32\pi F_\pi^2} \Bigl\{ 1+\epsilon + \Order(M_\pi^4) \Bigr\} ~, \no\\
\epsilon &=& \frac{5\Mpi}{84\pi^2 F_\pi^2} \left( {\bar\ell_1} + 2{\bar\ell_2} 
+ \frac{3}{8}{\bar\ell_3}
+ \frac{21}{10}{\bar\ell_4} + \frac{21}{8} \right) ~. \label{eq:a00p4}
\eea
There are two different types of LECs in the expression \eqref{eq:a00p4}.
$\bar \ell_1$, $\bar \ell_2$ come with structures containing four derivatives
(like $L_{1-3}$ in \eqref{eq:L4SU3}), i.e.\ they survive in the chiral limit
and can be determined from the momentum dependence of the $\pi\pi$ scattering amplitude,
namely from $d$-waves.
$\bar \ell_3$, $\bar \ell_4$ however are 
symmetry breaking terms that specify the quark mass dependence
(comparable to $L_{4,5}$ in \eqref{eq:L4SU3}), therefore they
cannot be determined from $\pi\pi$ scattering alone.

One additional observable to consider is the scalar form factor of the pion $\Gamma(s)$,
which is defined as
\beq 
\langle \pi^a(p) \pi^b(p') \,|\, \hat m(\bar u u+\bar d d)\,|\,0 \rangle
= \delta^{ab} {\Gamma(s)} ~, \quad s=(p+p')^2 ~.
\eeq
At tree level, one has  ${\Gamma(s)} = 2B\hat m = \Mpi + \Order(p^4) $
in accordance with the Feynman-Hellman theorem,
$
{\Gamma(0)} = \langle \pi \,|\, \hat m\, \bar qq\,|\,\pi\rangle = \hat m \,{\partial \Mpi}/{\partial\hat m} 
\,.
$
At next-to-leading order, one  defines the scalar radius $\langle r^2\rangle_\pi^S$ according to
\bea {\Gamma(s)} &=& \Gamma(0) \left\{ 1+\frac{1}{6}{\langle r^2\rangle_\pi^S} s + \Order(s^2) \right\}  
~, \no\\
{\langle r^2\rangle_\pi^S} &=& \frac{3}{8\pi^2 F_\pi^2} \left( {\bar\ell_4} - \frac{13}{12} \right) 
+ \Order(\Mpi) ~,
\eea
therefore the scalar radius is directly linked to $\bar \ell_4$.
Although the scalar form factor is not directly experimentally accessible,
one can analyse $\Gamma(s)$ in dispersion theory and extract $\langle r^2\rangle_\pi^S$
that way.

If we plug in the low-energy theorems and eliminate $\bar \ell_1$, $\bar \ell_2$, $\bar \ell_4$
in favour of the $d$-wave scattering lengths $a_2^{I=0,2}$ and the scalar radius, 
we can rewrite $\epsilon$ in \eqref{eq:a00p4} as
\beq
\epsilon = \frac{\Mpi}{3}{\langle r^2\rangle_\pi^S} 
+ \frac{200\pi}{7}F_\pi^2 \Mpi \left({a_2^0} + 2{a_2^2}\right)
- \frac{\Mpi}{672\pi^2 F_\pi^2} \left(15{\bar\ell_3}-353\right) ~.
\eeq
Therefore all we need to do is to measure $a_0^0$ and extract $\bar\ell_3$
from the above relation.  This will tell us 
how much of $\Mpi$ is due to the linear term in the quark mass expansion.


\subsection{Experiments on \boldmath{$\pi\pi$} scattering}

How can the $\pi\pi$ scattering lengths be measured?
We wish to briefly comment on four different possibilities to extract information
on the $\pi\pi$ interaction at low energies:
pion reactions on nucleons, in particular $\pi N \to \pi \pi N$;
so-called $K_{e4}$ decays $K^+ \to \pi^+ \pi^- e^+ \nu_e$ \cite{BNL-865,NA48-Ke4};
the cusp phenomenon in $K^+ \to \pi^+ \pi^0 \pi^0 $ \cite{K3pi-NA48};
and the lifetime of pionium~\cite{DIRAC}.

\subsubsection*{\boldmath{$\pi N \to \pi \pi N$}}

\begin{figure}
\begin{minipage}{6.2cm}
\begin{center}
\includegraphics[height=3cm]{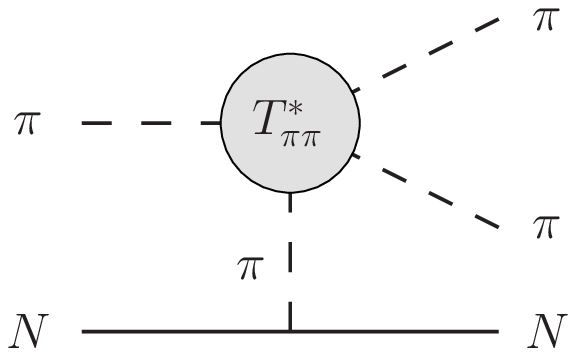}
\caption{One-pion exchange contributions to $\pi N \to \pi \pi N$.\label{fig:piNpipiN}}
\end{center}
\end{minipage}\hfill
\begin{minipage}{6.2cm}
\begin{center}
\includegraphics[height=3cm]{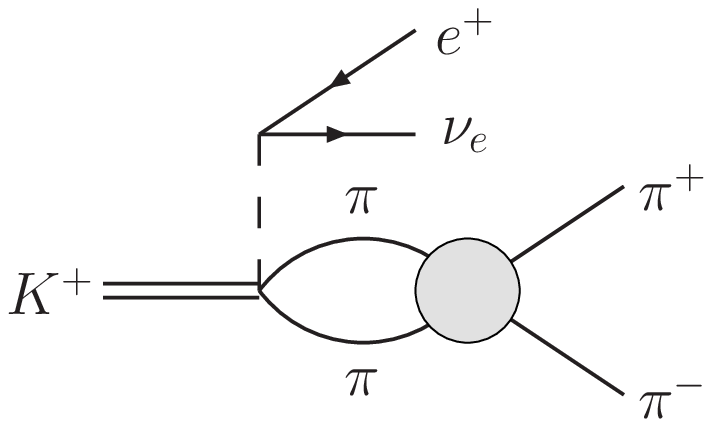}
\caption{$\pi\pi$ rescattering in $K_{e4}$ decays.\label{fig:Ke4}}
\end{center}
\end{minipage} 
\end{figure}
As shown schematically in Fig.~\ref{fig:piNpipiN}, the process $\pi N \to \pi \pi N$
receives contributions from graphs containing
the off-shell $\pi\pi$ amplitude $T_{\pi\pi}^*$, 
therefore it ought to be sensitive to $\pi\pi$ interactions.
In order to
isolate the one-pion-exchange contribution, one has to extrapolate to the pion pole at $t=\Mpi$,
which is outside the physical region; 
this procedure is not without difficulties, and the 
data obtained thereof tend to be at relatively high energies, 
so it does not give direct access to scattering lengths 
without further theoretical input;
see~\cite{BKMpipiN} for a ChPT-based analysis.

\subsubsection*{\boldmath{$K^+ \to \pi^+ \pi^- e^+ \nu_e$}}

Why is it possible to measure $\pi\pi$ scattering in such a kaon decay process?
The first, naive explanation, is that the  
pions undergo final-state interactions, see Fig.~\ref{fig:Ke4}, and they are 
the only strongly interacting particles in the final state, therefore
they ought to be sensitive to this interaction.
The more educated explanation is that
$K_{e4}$ decays can be described by form factors, 
which share the phases of $\pi\pi$ interaction due to Watson's final state theorem~\cite{Watson}.
It can be shown~\cite{PaisTreiman} that the interference between $s$- and $p$-waves can
be unambiguously extracted,
\[
\delta_0^0 (E_{\pi\pi}) - \delta_1^1 (E_{\pi\pi}) ~,
\]
and as $E_{\pi\pi}$ is kinematically restricted to be smaller than $M_K$,
these phases are measured close to threshold.

\subsubsection*{Cusp in \boldmath{$K^+ \to \pi^+ \pi^0 \pi^0 $}}

In a measurement of the kaon decay process $K^+ \to \pi^+ \pi^0 \pi^0 $,
the NA48/2 collaboration has detected a cusp phenomenon, i.e.\ a
sudden change in slope, in the  
invariant mass spectrum of the $\pi^0\pi^0$ pair
at$M_{\pi^0\pi^0}=2M_{\pi^+}$~\cite{K3pi-NA48}.
\begin{figure}
\begin{center}
\includegraphics[height=3cm]{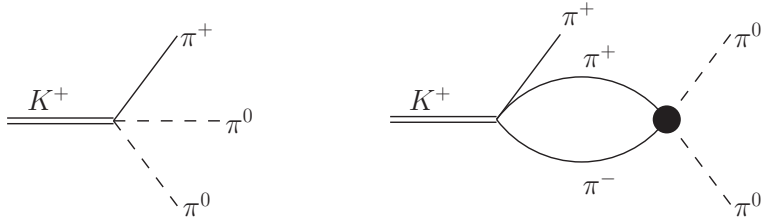}
\caption{``Direct'' and ``rescattering'' contributions to
$K^+ \to \pi^+ \pi^0 \pi^0 $. \label{fig:K3pi}}
\end{center} \vspace{-3mm}
\end{figure}
As suggested in Fig.~\ref{fig:K3pi}, this phenomenon 
can be explained by an interference effect between ``direct''
tree graphs for $K^+ \to \pi^+ \pi^0 \pi^0 $ and 
a $K^+ \to \pi^+ \pi^+ \pi^-$ decay, followed by $\pi^+\pi^-\to\pi^0\pi^0$ rescattering.
The one-loop graph has a smooth part plus a part $v_{+-}(s)$, where
\beq
v_{+-}(s) =
 \left\{
\begin{array}{ll}
-\frac{1}{16\pi}  \sqrt{\frac{4\Mpipm}{s}-1} ~, & s<4\Mpipm ~,\\[2mm]
\frac{i}{16\pi} \sqrt{1-\frac{4\Mpipm}{s}} ~, & s>4\Mpipm ~.
\end{array} 
\right.
\eeq
Below the $\pi^+\pi^-$ threshold, the loop graph is real and interferes 
directly with the tree contributions, while above threshold, it does not.
Due to the square-root behaviour of $v_{+-}(s)$, a cusp is seen~\cite{MMS,CI}.
The strength of this cusp is proportional to the scattering amplitude for
$\pi^+\pi^-\to\pi^0\pi^0$ at threshold, hence a combination of $\pi\pi$ scattering lengths.

This behaviour is complicated at two-loop order as
in contrast to $K_{e4}$, there are three strongly interacting particles in the final state
\cite{CI,CGKR,GPS}; 
in addition, virtual photons further modify the cusp structure.
Nevertheless, the high statistics available in the experimental data in principle
allow for a very precise determination of the scattering lengths, 
and the appropriate theoretical accuracy has to be provided.

\subsubsection*{Pionium lifetime}

Pionium is a hadronic atom, a $\pi^+\pi^-$ system bound by electromagnetism.
The energy levels of this system can in principle
be calculated as in quantum mechanics for the hydrogen atom,
however, they are perturbed by the strong interactions:
the ground state is not stable, it decays according to
\[
A_{\pi^+\pi^-} \to \pi^0\pi^0, \, \gamma\gamma,\,\ldots ~.
\]
The decay width is given by the following (improved) Deser formula~\cite{Deser,Gall99}
(further literature can be traced back from~\cite{GLR,Saz}):
\bea
\Gamma &=& \frac{2}{9} \alpha^3 p\, | {\mathcal{A}(\pi^+\pi^- \to \pi^0 \pi^0)_{\rm thr}}|^2 (1+\epsilon)\no\\
&=& \frac{2}{9} \alpha^3 p\, |{a_0^0 - a_0^2}|^2 (1+\delta) ~,
\eea
where $p$ is the momentum of the $\pi^0$ in the decay in the centre-of-mass frame, 
and $\epsilon$ and $\delta$ are numerical correction factors accounting for
isospin violation effects beyond leading order, $\delta = 0.058 \pm 0.012$~\cite{GLRG}.
Taking information on the scattering lengths from elsewhere, one can predict 
the pionium lifetime as
\beq \tau = (2.9 \pm 0.1) \times 10^{-15} s ~. 
\eeq
Ultimately, one however wants to turn the argument around, measure the lifetime 
and extract $a_0^0 -a_0^2$.  The corresponding experimental efforts
are undertaken by the DIRAC collaboration, first results have been obtained~\cite{DIRAC}.

\subsubsection*{Result on \boldmath{$\bar \ell_3$}} 

For reasons of brevity, we just compare 
to the BNL-865 result for $a_0^0$~\cite{BNL-865},
\beq 
a_0^0 = 0.216 \pm 0.013_{\rm stat} \pm 0.004_{\rm syst} \pm 0.005_{\rm theo} ~.
\label{eq:a00exp}
\eeq
For an up-to-date compilation of the various experimental results, 
see e.g.~\cite{Leutwyler06} and references therein.
From \eqref{eq:a00exp}, one can extract a value for $\bar \ell_3 \approx 6 \pm 10$,
which is compatible with the original estimate in~\cite{GL84}
as well as lattice determinations (see~\cite{ETM} and the discussion 
in~\cite{Leutwyler06}).
For the central value, the subleading correction for $M_\pi^2$ in \eqref{eq:Mpip4}
amounts to a mere 4\%, therefore
even from this seemingly rather loose bound, on can conclude that
the leading term in the quark mass expansion of the pion mass dominates by far~\cite{CGL01}.


\subsection{On the size of the corrections in \boldmath{$a_0^0$}}\label{sec:a00size}

If we remember the tree-level result for $a_0^0$~\eqref{eq:a0Op2},  
$a_0^0({\rm tree}) = 0.16$, the experimental result~\eqref{eq:a00exp} seems somewhat surprising:
higher-order corrections are of the order of $30\%$, 
rather than of the order of $\Mpi/(4\pi F_\pi)^2 \approx 2\%$.  
In order to understand this, we have to have another look at $\epsilon$ 
in \eqref{eq:a00p4}.
The $\bar \ell_i$ contain chiral logarithms, $\bar \ell_i \to - \log \Mpi $;
collecting these together, $\epsilon$ contains logarithmic terms 
\beq 
\epsilon ~=~ - \frac{9\Mpi}{32\pi^2 F_\pi^2} {\log\frac{\Mpi}{\mu^2}} 
\eeq
which, estimated at a scale $\mu=1$~GeV, alone amount to 25\%.
We conclude that
chiral logarithms potentially enhance higher-order corrections,
and that the isoscalar $s$-wave $\pi\pi$ scattering length contains chiral logarithms 
with rather large coefficients, therefore corrections to the tree-level result
are sizeable.


\subsection{Quark mass ratios revisited}

As another application of ChPT beyond leading order, we want to briefly
revisit the ratios of the light quark masses.
Forming dimensionless ratios, it turns out 
that one can write the $\Order(p^4)$ corrections in the form~\cite{GL85}
\bea
\frac{\MK}{\Mpi} &=& \frac{m_s+\hat m}{m_u+m_d} \Bigl\{ 1 + {\Delta_M} + \Order(m_q^2) \Bigr\} ~,\no\\
\frac{(\MKn-\MKp)_{\rm strong}}{\MK - \Mpi} &=& 
\frac{m_d-m_u}{m_s-\hat m} \Bigl\{ 1 +{ \Delta_M} + \Order(m_q^2) \Bigr\} ~,
\no \\
{\rm where} \quad {\Delta_M} &=& \frac{8(\MK-\Mpi)}{F_\pi^2} (2L_8-L_5) + {\rm chiral~logs} ~.\label{eq:DeltaM}
\eea 
The double ratio $Q^2$ is therefore particularly stable with respect to higher-order corrections,
\beq
Q^2 = \frac{m_s^2-\hat m^2}{m_d^2-m_u^2} = \frac{\MK}{\Mpi} \frac{\MK-\Mpi}{(\MKn-\MKp)_{\rm strong}}
\Bigl\{ 1 + \Order(m_q^2) \Bigr\} ~. \label{eq:Q2}
\eeq
\eqref{eq:Q2} can be rewritten in the form of an ellipse equation for the quark mass ratios
$m_u/m_d$, $m_s/m_d$ (Leutwyler's ellipse~\cite{Leutwyler96}),
\beq 
\left(\frac{m_u}{m_d}\right)^2 + \frac{1}{{Q^2}} \left(\frac{m_s}{m_d}\right)^2 = 1 ~.
\eeq
We can use Dashen's theorem \eqref{eq:Dashen} to determine $(\MKn-\MKp)_{\rm strong}$
and therefore $Q$, with the result
\beq
Q_{\rm Dashen} = 24.2 ~.
\eeq
However, corrections to Dashen's theorem of $\Order(e^2m_q)$ 
are potentially large, different models yield a range~\cite{DashenCorr}
\beq 
1 ~\lesssim~ (\MKp - \MKn)_{\rm em}/ (\Mpip - \Mpin)_{\rm em} ~\lesssim~ 2.5 ~,
\eeq
inducing a rather large uncertainty in $Q$,
$20.6 \lesssim Q \lesssim 24.2$.
It would therefore be most desirable to obtain information on $Q$
independent on meson mass relations. One such source 
will be introduced in the following subsection.


\subsection{\boldmath{$\eta \to 3\pi$}}

The $\eta$ meson has isospin $I=0$, while
three pions with angular momentum 0 cannot have $I=0$, but only $I=1$.
$\eta \to 3\pi$ is therefore an isospin violating decay.
Using the leading chiral Lagrangian $\Lagr^{(2)}$ (including isospin breaking), 
the tree amplitude for $\eta \to \pi^+ \pi^- \pi^0$ can be calculated to be
\beq
A(s,t,u) = \frac{B({m_u-m_d})}{3\sqrt{3}F_\pi^2} \left\{ 1 + \frac{3(s-s_0)}{\Me-\Mpi} \right\} ~,
\label{eq:eta3pi}
\eeq
where $s=(p_\eta-p_{\pi^0})^2$, $t=(p_\eta-p_{\pi^+})^2$, $u=(p_\eta-p_{\pi^-})^2$, 
$3s_0 = s+t+u = \Me + 2\Mpip + \Mpin$.
For $\eta \to 3\pi^0$, one finds the amplitude $A(s,t,u)+A(t,u,s) + A(u,s,t)$,
with $A$ as given in \eqref{eq:eta3pi}.
We note that the
electromagnetic term \eqref{eq:L2em} does not contribute~\cite{Sutherland}.
Terms of order {$e^2 m_q$} were found to be very small~\cite{Baur95},
therefore $\eta \to 3\pi$  (potentially) allows for a much cleaner access to $m_u-m_d$
than the meson masses.
We can rewrite $A(s,t,u)$ explicitly in terms of $Q^2$,
\bea
A(s,t,u) &=& \frac{1}{{Q^2}} \frac{\MK}{\Mpi} \left(\Mpi-\MK\right) \frac{M(s,t,u)}{3\sqrt{3}F_\pi^2} ~,\no\\
M(s,t,u) &=& \frac{3s-4\Mpi}{\Me-\Mpi} \quad{\rm (at~leading~order)\,.}
\eea
Problems here arise from the fact
that there are strong final-state interactions among the three pions:
the one-loop corrections increase the width by a factor of 2.5, 
\bea
\textrm{tree:} \quad \Gamma(\eta \to \pi^+\pi^-\pi^0) &=& 66~{\rm eV} ~,\no \\
\textrm{one-loop:} \quad \Gamma(\eta \to \pi^+\pi^-\pi^0) &=& 160\pm50~{\rm eV} \quad\cite{GLeta}~, 
\eea
and even higher-order corrections are not negligible.
Furthermore, there are partially contradictory experimental results 
(in particular on Dalitz plot parameters).
This is therefore still a process of current interest, 
with strong ongoing experimental efforts for both final states~\cite{eta3piExp}.

\begin{figure}
\begin{center}
\includegraphics[width=0.7\linewidth]{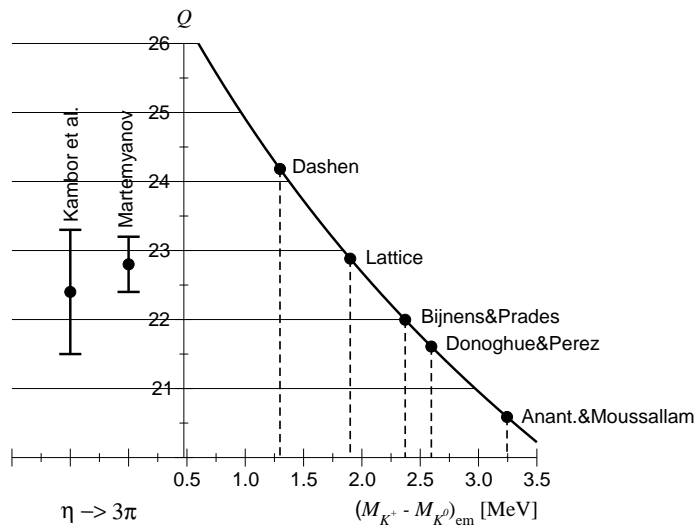}
\caption{Various results on $Q$~\cite{DashenCorr,kambor,eta3pi}.  
The figure is inspired by \cite{Leutwyler96}.\label{fig:Qplot}}
\end{center} 
\end{figure}

The combined information on $Q$ as deduced from various
corrections to Dashen's theorem is shown in Fig.~\ref{fig:Qplot}, 
together with two results obtained from studies of $\eta \to 3\pi$~\cite{kambor,eta3pi}
(see also the discussion in~\cite{borasoynissler}).
Even with $Q$ fixed, additional constraints are 
needed to find the position on the ellipse.
Two possibilities are
information on $\eta \eta'$ mixing, which, together with large--$N_c$ arguments, 
leads to a determination of $\Delta_M$ in \eqref{eq:DeltaM}.
Alternatively, the ratio $R = (m_s-\hat m)/(m_d-m_u)$ can be extracted 
from baryon masses~\cite{GLPhysRept}.


\section{Chiral perturbation theory with baryons}\label{sec:baryonChPT}

So far, we have considered an effective field theory exclusively for (pseudo) Goldstone bosons.
Perhaps the most important extension of this theory is the inclusion
of nucleons or the baryon ground state octet in chiral SU(2) and SU(3), respectively.
The problem, as we shall see later, is that the
nucleon mass is a new, heavy mass scale that does not vanish in the chiral limit,
\[ {\lim_{m_q\to 0} m_N \approx m_N } ~.\]
The idea for their incorporation in the theory is to view nucleons as (massive) matter fields 
coupled to pions and external sources.
Their 3-momenta ought to remain small, of the order of $M_\pi$, in all processes.
The number of baryons is therefore conserved, we consider
no baryon--antibaryon creation/annihilation.
In particular, in these lectures, we confine ourselves to processes with exactly one baryon.

For the construction of the meson-baryon Lagrangian, we proceed as before:
we choose a suitable representation and 
transformation law for baryons under ${\rm SU}(N)_L \times {\rm SU}(N)_R$,  
and organise the effective Lagrangian according to an increasing number of momenta.

It turns out to be convenient to introduce a new field $u$ for the Goldstone boson fields
according to $u^2=U$, which transforms as
\beq
u ~\longmapsto~ \sqrt{LUR^\dagger} = L \,u\, {K^\dagger(L,R,U)} 
= {K(L,R,U)} \,u\, R^\dagger  ~.\label{eq:utrafo}
\eeq
Here, ${K(L,R,U)} \in {\rm SU}(N)$ is the so-called compensator field, 
that depends in a non-trivial way on $L$, $R$, and $U$.
For SU$(N)_V$ transformations ($L=R$), \eqref{eq:utrafo} obviously reduces to $K(L,R,U)=L=R$.

A particularly convenient representation for nucleons and baryons is given by the following:
\bea
\psi = \biggl(\!\begin{array}{c}p\\n\end{array}\!\biggr) &\longmapsto&  \,{\psi' = K \psi} ~,\no
\\
B = \left(\! \begin{array}{ccc} 
\frac{\Sigma^0}{\sqrt{2}}+\frac{\Lambda}{\sqrt{6}} & \Sigma^+ & p\\
\Sigma^- & -\frac{\Sigma^0}{\sqrt{2}}+\frac{\Lambda}{\sqrt{6}} & n\\
\Xi^-  & \Xi^0 & -\frac{2\Lambda}{\sqrt{6}} \end{array}\!\right)
&\longmapsto& {B' = K B K^{-1}}  ~.
\eea
We introduce a covariant derivative
$ D^\mu = \partial^\mu + \Gamma^\mu$
with the chiral connection $\Gamma^\mu$ (vector)
\beq
\Gamma^\mu = \frac{1}{2} \bigl( u^\dagger (\partial^\mu-i\,r^\mu)u + u(\partial^\mu-i\,l^\mu)u^\dagger\bigr) 
\eeq
that transforms according to
$ \Gamma^\mu \mapsto K \Gamma^\mu K^\dagger - (\partial^\mu K) K^\dagger$, 
such that the covariant derivative has the expected transformation behaviour
\[ D^\mu \psi \longmapsto K D^\mu \psi ~.\]
Furthermore, we shall use the chiral vielbein (axial vector) 
\beq 
u^\mu = i \bigl( u^\dagger (\partial^\mu-i\,r^\mu)u - u(\partial^\mu -i\,l^\mu)u^\dagger\bigr) 
\eeq
that transforms according to
$ u^\mu \mapsto K u^\mu K^\dagger$.
Finally, we can rewrite the (pseudo)scalar source term $\chi = 2B(s+i\,p) = 2B{\M}+\ldots$ as
\beq \chi_+ = u^\dagger \chi u^\dagger + u \chi^\dagger u ~, \eeq
such that $\chi_+ \mapsto K \chi_+ K^\dagger $, and all the constitutive elements
of the chiral Lagrangian transform with the compensator field $K$.


\subsection{The leading-order chiral meson-baryon Lagrangian}

We are now in the position to write down the leading-order meson-baryon Lagrangian, both
for chiral SU(2) and SU(3):
\bea
\Lagr^{(1)}_{\pi N} \!\!&=&\!\! \bar\psi \left( i\gamma_\mu D^\mu -{m} +\frac{{g_A}}{2} \gamma_\mu \gamma_5 u^\mu 
\right) \psi ~, \label{eq:LpiN1}\\
\Lagr^{(1)}_{\phi B} \!\!&=&\!\! \langle \bar B \left(i\gamma_\mu D^\mu -{m}\right) B\rangle 
+ \frac{D/F}{2} \langle \bar B\gamma_\mu\gamma_5 [u^\mu,B]_\pm \rangle ~.\label{eq:LphiB1}
\eea
While in meson ChPT, the Lagrangians come only in even powers of derivatives or momenta, 
odd powers of momenta are allowed in the meson-baryon sector due to 
the presence of spin (or, more general, Dirac structures):
\[ \Lagr_{\pi N} =  \Lagr_{\pi N}^{(1)} +  \Lagr_{\pi N}^{(2)}+  \Lagr_{\pi N}^{(3)}+  \Lagr_{\pi N}^{(4)} + \ldots ~.\]
The new parameters of $\Lagr^{(1)}_{\pi N}$ comprise
$m$, the nucleon (baryon) mass in the chiral limit;
$g_A$ (in SU(2)), which, upon expansion of $u_\mu = 2 a_\mu + \Order(\pi)$,
can be identified with the 
axial vector coupling that is known from neutron beta decay, $g_A = 1.26$;
or $D / F$, the two axial vector couplings in SU(3), which can be determined
from semileptonic hyperon decays and have to fulfil the
SU(2) constraint $D+F=g_A$ ($D\approx 0.80, F\approx 0.46$).


\subsection{Goldberger--Treiman relation}

As a first consequence of \eqref{eq:LpiN1}, we want to derive the so-called
Goldberger--Treiman relation.
Setting external sources to zero, $v_\mu = a_\mu=0$, and expanding in powers 
of the pion field, we find 
$u_\mu = -\partial_\mu\pi/F_\pi + \Order(\pi^3)$
for the chiral vielbein, 
which results in the $\pi NN$ vertex 
\beq \Lagr^{(1)}_{\pi N} \to - \frac{g_A}{2F_\pi} {\bar \psi} \gamma^\mu \gamma_5 {\partial_\mu \pi \psi} ~.
\eeq
The corresponding Feynman rule looks as follows:
\begin{center}
\begin{tabular}{rl}
\begin{minipage}{3cm}
\includegraphics[width=2.5cm]{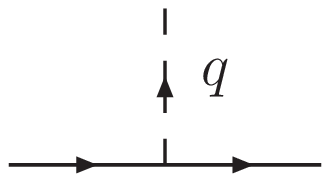}
\end{minipage} &
$ \dfrac{g_A}{2F_\pi} \sh q \gamma_5 \vec \tau ~\doteq~ V_{\pi NN} ~.$
\end{tabular}
\end{center}
We can deduce a  $N \to \pi N$ transition amplitude 
\[ T_{\pi NN} = -i \bar u(p') V_{\pi NN} u(p) = -i {\frac{g_A m_N}{F_\pi}} \bar u(p') \gamma_5 u(p) \vec \tau ~,\]
which, compared to the canonical amplitude $-i\, {g_{\pi N}} \bar u(p') \gamma_5 u(p) \vec \tau $,
yields the relation
\beq
g_{\pi N} ~=~ \dfrac{g_A m_N}{F_\pi} ~. \label{eq:GT}
\eeq
\eqref{eq:GT} is 
remarkable for relating weak ($g_A$) and strong ($F_\pi$) interaction quantities.
Numerically it is rather well fulfilled in nature, $13.1 \ldots 13.4 = 12.8$.


\subsection{Weinberg's power counting for the one-baryon sector}

We can derive a similar power counting formula as in \eqref{eq:powercounting}
for the one-baryon sector.
The chiral dimension $\nu$ of an arbitrary $L$-loop diagram 
with $V_d^{\pi\pi}$ meson--meson vertices of order $d$
and  $V_{d'}^{\pi N}$ meson--baryon vertices of order $d'$ is given by \vspace{-1mm}
\beq
 \nu = 2L+1 + \sum_d V_d^{\pi\pi}(d-2) + \sum_{d'} V_{d'}^{\pi N}(d'-1) ~. \label{eq:powercountingpiN}
\eeq
Note that  $d\geq 2$, $d'\geq 1$, such that the right hand side of
\eqref{eq:powercountingpiN} again consists of a sum of non-negative terms only.
We conclude that at
$\Order(p^1)$, $\Order(p^2)$, only tree diagrams contribute; 
tree plus one-loop diagrams enter at 
$\Order(p^3)$, $\Order(p^4)$;
tree, one-loop, and two-loop diagrams can contribute to
$\Order(p^5)$, $\Order(p^6)$, etc.

However, as it was noted in~\cite{GSS}, in contrast to the meson sector, 
loop graphs do not necessarily obey the naive power counting rules
as put forward in \eqref{eq:powercountingpiN}.
The reason is that loop integrals cover all energy scales:
while in the Goldstone boson sector, all mass scales are ``small'',
such that naive power counting has to work in a mass-independent regularisation
scheme (like dimensional regularisation),
there is a new mass scale in the nucleon sector, the nucleon mass
$m_N \approx \Lambda_\chi \approx 1$~GeV.  The loop integration
then also picks up momenta $p \sim m_N$.

\begin{figure}
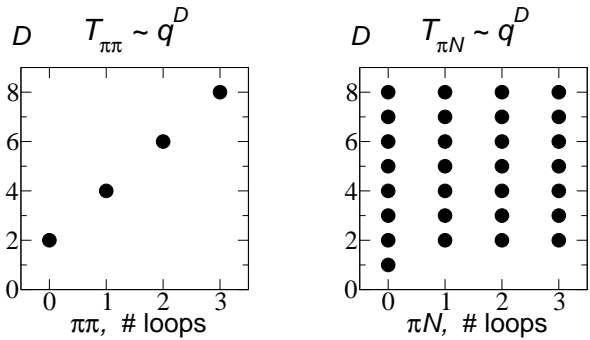

\begin{center}
\includegraphics[width=0.25\linewidth]{powerCounting_pipi} \hspace{1cm}
\includegraphics[width=0.25\linewidth]{powerCounting_piN}
\caption{Schematic representation of loop contributions
to chiral amplitudes in the Goldstone boson (left panel)
and the meson-baryon (right panel) sector.  
The figure is inspired by~\cite{GSS}. \label{fig:GSS}}
\end{center} \vspace{-3mm}
\end{figure}

Schematically, the situation is depicted in Fig.~\ref{fig:GSS}:  
in contrast to the Goldstone boson sector, higher-order loop graphs 
in the meson-baryon theory renormalise lower-order couplings
at each order in the loop expansion.
In the following, we shall discuss two possible remedies to restore
the features of naive power counting for meson-baryon ChPT.


\subsection{Heavy-baryon ChPT}

The first remedy goes by the name of heavy-baryon chiral 
perturbation theory (HBChPT) \cite{JenkinsManohar,BKKM}.
It is constructed in close analogy to heavy-quark effective field theory:
we decompose the baryon momentum into a large part
proportional to the nucleon velocity, plus a small residual momentum
according to
\beq p_\mu ~=~ m_N v_\mu + l_\mu ~, \quad
v^2=1 ~, \quad v\cdot l \ll m_N ~.
\eeq
In the heavy-baryon limit, the nucleon propagator is then of the form
\[
\frac{1}{p^2-m_N^2} ~\to~ \frac{1}{2m_N} \frac{1}{v\cdot l} + \Order\bigl(1/m_N^2\bigr) ~.
\]
This procedure eliminates the mass scale {$m_N$} from the propagator, which re-enters as a
parametrical suppression factor.
HBChPT is then a two-fold expansion in powers of $(p/\Lambda_\chi)^n$ and
$(p/m_N)^n$, where $m_N \approx \Lambda_\chi$.

The nucleon field $\psi$ is decomposed into velocity eigenstates according to
\beq
H_v(x) = e^{im_N v\cdot x} P_v^+ \psi(x) ~, \quad
h_v(x) = e^{im_N v\cdot x} P_v^- \psi(x) ~,
\eeq
where we have used the projectors $P_v^\pm = \frac{1}{2}(1\pm\!\Sh v) $
onto velocity eigenstates.  $H_v$ represents the ``big'' components 
of the spinor at low energies, while $h_v$ are the ``small'' components.
The exponential eliminates the large mass term from the time evolution
of the field $H_v$.
Written in terms of the new field $H_v$, 
the leading-order Lagrangian $\Lagr_{\pi N}^{(1)}$ becomes 
\beq 
\Lagr_{\pi N}^{(1)} = \bar H_v \bigl( i v\cdot D + g_A { S} \cdot u \bigr) H_v + \Order\bigl({1/m_N}\bigr) ~,
\label{eq:LpiN1HB}
\eeq
where we have introduced the Pauli-Lubanski spin vector ${S_\mu} = \frac{i}{2}\gamma_5 \sigma_{\mu\nu}v^\nu$.
We note that the nucleon mass does not occur in \eqref{eq:LpiN1HB}, and the 
Dirac structure is massively simplified.
$1/m_N$ corrections can be constructed systematically on the Lagrangian level in analogy to
Foldy--Wouthuysen transformations~\cite{BKKM}.


\subsection{Infrared regularisation}

The second procedure is called infrared regularisation~\cite{Becher} (see also~\cite{ET,LP,MainzIR}
for variants of this approach).
It is an alternative way to regularise loop integrals and allows for
a manifestly covariant way to calculate loops in baryon ChPT.

\begin{figure}
\begin{center}
\includegraphics[height=1.2cm]{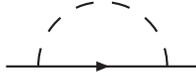} 
\caption{Nucleon self-energy graph. Full and dashed lines denote nucleon and pion,
respectively.\label{fig:piNloop}}
\end{center} \vspace{-3mm}
\end{figure}

Let us consider the (relativistic) nucleon self-energy graph in Fig.~\ref{fig:piNloop}.
Evaluated at threshold in $d$ dimensions, it yields the result
\beq
\frac{\Gamma(2-\frac{d}{2})}{(4\pi)^{d/2}(d-3)} \frac{{m_N^{d-3}}+{M_\pi^{d-3}}}{m_N+M_\pi} ~.
\label{eq:piNloop}
\eeq
We decompose \eqref{eq:piNloop} into two parts, the ``regular'' part $\propto m_N^{d-3}$, 
and the ``infrared'' part $\propto M_\pi^{d-3}$.
The following properties of this decomposition can be shown to hold in general.
The regular part scales with fractional powers of $m_N$, but has a regular
expansion in $M_\pi$ and momenta.  It is this part that violates naive power counting,
but as it can always be expanded as a polynomial, it can be absorbed by a redefinition
of contact terms in the Lagrangian.  The infrared part, on the other hand, 
scales with fractional powers of $M_\pi$; it contains all the ``interesting'' pieces
of the loop diagram such as non-analytic structures and imaginary parts, and it obeys
the naive power counting rules.  It is therefore only the infrared part of the
loop integral that we want to retain.

In~\cite{Becher}, a very simple prescription to isolate the infrared part
of the loop integral was given.  
With $a = \Mpi - k^2 -i\epsilon$, $b = m_N^2 - (P-k)^2 - i\epsilon$, 
the scalar loop integral corresponding to Fig.~\ref{fig:piNloop} can be written as
\beq
H = \frac{1}{i}\int \frac{d^dk}{(2\pi)^d} \frac{1}{ab}
= {\int_0^1 dz} \frac{1}{i}\int \frac{d^dk}{(2\pi)^d} \frac{1}{[{(1-z)}a+ zb]^2} ~,
\label{eq:FeynPar}
\eeq
where we have introduced the Feynman parameter $z$.
The Landau equations can be used to analyse the singularity structure of \eqref{eq:FeynPar}
in terms of values of $z$ and the kinematic position where they occur:
\begin{center}
\begin{tabular}{llcl}
{$z=\frac{M_\pi}{m_N+M_\pi}$} &leading (pinch) singularity &$\leftrightarrow$ & {$P^2 = (m_N+M_\pi)^2$}~, \\[1.5mm]
{$z=0$} & endpoint singularity &$\leftrightarrow$ &{$\Mpi=0$} ~,\\[1.5mm]
{$z=1$} & endpoint singularity &$\leftrightarrow$ &{$m_N^2=0$} ~.\\[1mm]
\end{tabular}
\end{center}
Clearly, $z=1 \leftrightarrow m_N^2=0$ is not a ``low-energy'' singularity,
hence one can obtain the infrared part $I$ of $H$ by avoiding the $z=1$ endpoint singularity
and extending the integration to $z=\infty$:
\beq 
I = 
\int_0^\infty dz \frac{1}{i}\int \frac{d^dk}{(2\pi)^d} \frac{1}{[{(1-z)}a+ zb]^2}
~.
\eeq

\begin{figure}
\begin{center}
\includegraphics[width=\linewidth]{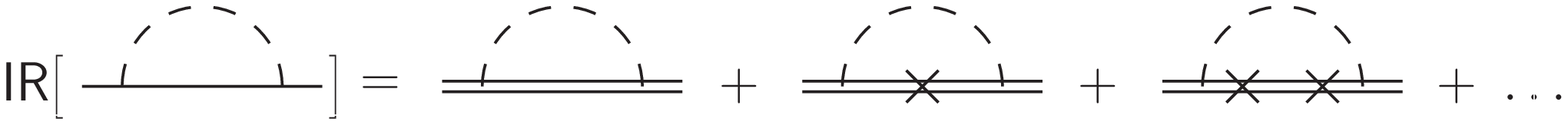}
\caption{Schematic relation of infrared regularisation and heavy-baryon expansion.
The double line denotes the heavy-baryon propagator, crosses denote 
$1/m_N$ insertions.\label{fig:IR-HB}}
\end{center} \vspace{-5mm}
\end{figure}
The relation between the infrared regularisation scheme and the heavy-baryon expansion
is shown schematically in Fig.~\ref{fig:IR-HB}: 
the infrared part $I$ corresponds to a 
resummation of all $1/m_N$ corrections in a certain heavy-baryon diagram.
This seems to be just the reverse of the expansion used to obtain
the heavy-baryon propagator in the first place; however, we have interchanged
summation and (highly irregular) loop integration, and the difference
in the order of taking these limits is the regular part of the loop integral.

In general, heavy-baryon loop integrals are easier to perform explicitly 
than those in infrared regularisation (although keeping track of all possible
$1/m_N$ corrections is a considerable task in HBChPT when going to subleading
loop orders).  So does the difference between both procedures ever matter?
Is the additional effort to calculate baryon ChPT in a manifestly Lorentz-invariant
fashion worth it?

\begin{figure}
\begin{center}
\includegraphics[width=3.5cm]{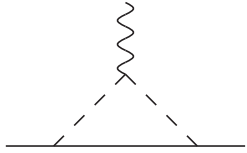}
\caption{Triangle graph contributing to the electromagnetic form factors of the nucleon.\label{fig:triangle}}
\end{center} \vspace{-3mm}
\end{figure}

Sometimes, the difference does indeed matter, and in order to see this,
we consider the electromagnetic form factors of the nucleon, defined by
\beq 
\langle N(p') | J_\mu^{em} | N(p) \rangle = e \bar u(p') \Bigl\{
\gamma_\mu F_1^N(t) + \frac{i\sigma_{\mu\nu}q^\nu}{2m_N} F_2^N(t) \Bigr\} u(p) ~.
\eeq
An important contribution to the spectral function ${\rm Im}\,F_1^v(t)$
(where $F_1^v = F_1^p-F_1^n$) stems from the so-called triangle graph, 
Fig.~\ref{fig:triangle}.
The ``normal'' threshold of the spectral function is at $t_{\rm thr}=4\Mpi$; however
there is an anomalous threshold very close by at 
$t_{\rm anom}=4\Mpi-M_\pi^4/m_N^2$.
As both coincide in the heavy-baryon limit, $t_{\rm anom} = 4\Mpi + \Order(M_\pi^4)$, 
the analytic structure is distorted, 
as we can see comparing the contributions to the spectral function
in infrared regularisation~\cite{Kubis00},
\beq
{\rm Im}\,F_1^v(t) ~\stackrel{{\rm IR}}{=}~ \frac{g_A^2}{192\pi F_\pi^2}(4m_N^2-\Mpi) \left(1-\frac{4\Mpi}{t}\right)^{{3/2}} + \ldots ~, 
\eeq
and in HBChPT~\cite{BKMspectral},
\beq
{\rm Im}\,F_1^v(t) ~\stackrel{{\rm HB}}{=}~ \frac{g_A^2}{96\pi F_\pi^2}(5t-8\Mpi) \left(1-\frac{4\Mpi}{t}\right)^{{1/2}} + \ldots  ~.
\eeq
While the infrared spectral function has the expected $p$-wave characteristic
$\propto (1-4\Mpi/t)^{3/2}$, the threshold behaviour of the heavy-baryon spectral function
is distorted.

\begin{figure}
\begin{center}
\includegraphics[width=8cm]{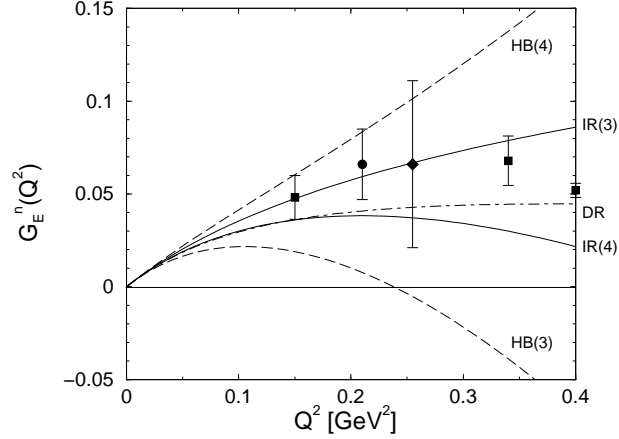}
\caption{Neutron electric form factor at third and fourth order, in HBChPT and infrared regularisation,
as a function of $Q^2=-t$.
Figure taken from~\cite{Kubis00}.\label{fig:GEn}}
\end{center} \vspace{-3mm}
\end{figure}

Furthermore, the resummation of relativistic recoil effects in the nucleon propagator
sometimes also helps to improve the phenomenological description of certain observables, 
as can be seen for the example of the neutron electric form factor
\beq 
G_E^n(t) = F_1^n(t) + \frac{t}{4m_N^2} F_2^n(t) ~,
\eeq
in Fig.~\ref{fig:GEn}.
While convergence between leading and subleading loop order ($\Order(p^3)$ and $\Order(p^4)$, 
respectively) in HBChPT is rather poor, it is improved in infrared regularisation 
and, in addition, much closer to the data.


\subsection{Quark mass dependence of the nucleon mass}

As a further application of ChPT for nucleons, let us consider
the quark mass expansion of the nucleon mass up to $\Order(p^3)$.
The result is
\beq
m_N = m - {4c_1 \Mpi} - {\frac{3g_A^2 M_\pi^3}{32\pi F_\pi^2}} + \Order(M_\pi^4) ~.\label{eq:mNOp3}
\eeq
The pion loop graph Fig.~\ref{fig:piNloop} yields a non-analytic term $\propto M_\pi^3 \propto \hat m^{3/2}$,
but the leading correction term comes from a counterterm in 
$\Lagr_{\pi N}^{(2)} = c_1 \bar\psi \langle\chi_+\rangle\psi + \ldots\,$,
with an unknown low-energy constant $c_1$.
$c_1$ is closely related to the so-called $\pi N$ $\sigma$-term.
We define the scalar form factor of the nucleon according to
\beq
\langle N(p') | \hat m(\bar u u+\bar d d)| N(p)\rangle = \sigma(t) \bar u(p')u(p) ~, \quad
t=(p-p')^2 ~.
\eeq
The $\sigma$-term is then given by
\beq
\sigma \equiv \sigma(0) = \frac{\hat m}{2m_N} \langle N | \bar u u + \bar d d | N \rangle ~.
\eeq
$\sigma$ to $\Order(p^3)$ can be calculated from \eqref{eq:mNOp3} 
using the Feynman--Hellman theorem,
\beq
\sigma = \hat m \dfrac{\partial m_N}{\partial \hat m} = 
-{4c_1 \Mpi} - {\dfrac{9g_A^2 M_\pi^4}{32\pi F_\pi^2}} + \Order(M_\pi^4) ~,
\eeq
so is is indeed given by $c_1$ at leading order.
Furthermore, $\sigma$ is related to the interesting question:
how much do strange quarks contribute to nucleon properties?
In this case, we consider the strangeness contribution to $m_N$, which is 
related to $\sigma$ by
\beq
\sigma = \frac{\hat m}{2m_N} \frac{\langle N| \bar uu + \bar dd -2\bar ss|N\rangle}{1-{y}} ~, \quad
y = \frac{2\langle N|\bar ss|N\rangle}{\langle N|\bar uu+\bar dd|N\rangle} ~.
\eeq
Now $(m_s-\hat m)(\bar uu+\bar dd-2\bar ss)$ is the part of $\Lagr_{\rm QCD}$ that produces the
SU(3) mass splittings and can therefore be related to differences in the baryon octet masses,
\beq \sigma = \frac{{\hat \sigma}}{1-{y}} ~, \quad
\hat \sigma = \frac{\hat m}{m_s-\hat m} \bigl( m_\Xi+m_\Sigma-2m_N\bigr) \simeq 26~{\rm MeV} ~.
\label{eq:sigmay}
\eeq
Higher-order corrections lead to a modified value $\hat \sigma \to (36 \pm 7)$~MeV~\cite{Borasoy}
(see also references therein for background, e.g.~\cite{GasserSigma,BKMmasses}).
But we conclude that,
if we know $\sigma$, we know $y$ and can therefore deduce information on the
strangeness content of the nucleon.

We remember that we learnt about the quark mass dependence of $\Mpi$ from $\pi\pi$ scattering;
it turns out that, here again, the quark mass dependence of $m_N$ and the $\sigma$-term
are closely related to $\pi N$ scattering.
$\pi N$ scattering amplitudes can be separated into 
isospin even and odd parts, 
\beq 
T_{\pi N}^\pm = \frac{1}{2} \bigl[
T(\pi^- p \to \pi^- p) \pm T(\pi^+ p \to \pi^+ p) \bigr] ~,
\eeq
and we can decompose $T^\pm$ further into spin flip/non-flip amplitudes.
Without going into all the details, 
let us consider the specific combinations $\bar D^\pm$ of $\pi N$ amplitudes.
In ChPT, the following relation can be proven:
\beq
\Sigma \equiv F_\pi^2 \bar D^+ \bigl(s=u=m_N^2,t=2\Mpi\bigr) 
= \sigma(2\Mpi) + \Delta_R ~. \label{eq:sigmaCD}
\eeq
The specific kinematic point $s=u=m_N^2,\,t=2\Mpi$ at which $\bar D^+$ is to be evaluated
is known as the Cheng--Dashen point.
\begin{figure}
\begin{center}
\includegraphics[width=8cm]{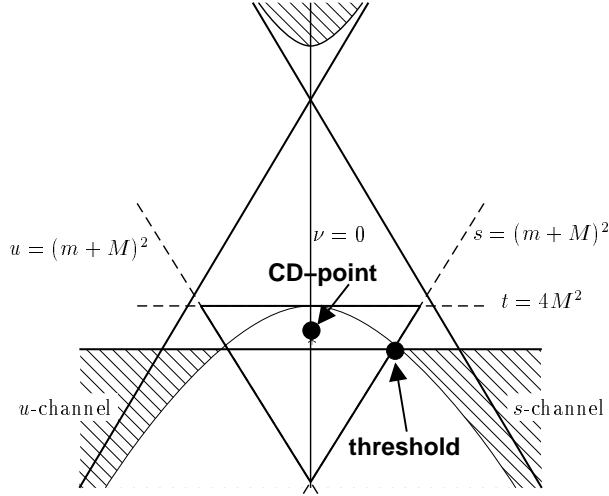}
\caption{Mandelstam plane for $\pi N$ scattering, with the position
of the $s$-channel threshold and the Cheng--Dashen (CD) point.  The figure is
a modified version of one taken from~\cite{Buettiker}.\label{fig:mandel}}
\end{center} \vspace{-3mm}
\end{figure}
As shown in Fig.~\ref{fig:mandel}, it lies in the unphysical region
(which is, in the $s$-channel, limited by $s\geq (m_N+M_\pi)^2$, $t\leq 0$).
The remainder
$\Delta_R = \Order(M_\pi^4)$ in \eqref{eq:sigmaCD} is very small, 
$\Delta_R \lesssim 2$~MeV~\cite{BKMsigma}.

The procedure to learn about the strangeness content of the nucleon therefore consists
of the following steps: (1) from $\pi N$ scattering, deduce the amplitude $\bar D^+$ 
at the Cheng--Dashen point; (2) use \eqref{eq:sigmaCD} to calculate $\sigma(2\Mpi)$;
(3) extrapolate $\sigma(t)$ to $t=0$; (4) calculate $y$ from $\sigma$ using \eqref{eq:sigmay}.

Step (1), the extrapolation into the unphysical region, is done using dispersion relations,
with the result
\beq
\Sigma = 60 \pm 7~{\rm MeV} ~.
\eeq
Step (2) is then safe due to the smallness of $\Delta_R$.
The crucial part is step (3),
\beq
\sigma(2\Mpi) = \sigma(0) + \Delta\sigma ~,
\eeq
i.e.\ we have to understand the $t$-dependence of the scalar form factor $\sigma(t)$.
A crude estimate would be to linearise the form factor,
\beq
\sigma(t) = \sigma(0) \Bigl\{ 1 + \dfrac{1}{6}{\langle r^2\rangle_\sigma} \,t + \ldots \Bigr\} ~,
\eeq
and assume $\langle r^2\rangle_\sigma \simeq \langle r^2\rangle_{\rm EM} = 0.8~{\rm fm}^2 $,  
leading to $\Delta \sigma \approx 3.5~{\rm MeV}$.
A complete one-loop calculation yields a value not too far from this,  $\Delta \sigma = 4.6$~MeV.
We would deduce
$\sigma \approx 55~{\rm MeV}$ and, consequently,
\beq
y=1-\dfrac{\hat \sigma}{\sigma} = 1-\dfrac{35~{\rm MeV}}{55~{\rm MeV}} \approx 0.4 ~,
\eeq
which appears far bigger than what one would naively expect.
This was for some time known as the ``$ \sigma$-term puzzle'':
are 300~MeV of nucleon mass due to strange quarks?

\begin{figure}
\begin{center}
\includegraphics[width=5cm]{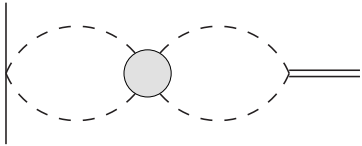}
\caption{The scalar form factor of the nucleon beyond one loop.
Full, dashed, and double lines denote nucleons, pions, 
and the scalar source, respectively.\label{fig:scalarFF}}
\end{center} \vspace{-3mm}
\end{figure}

The resolution lies in step (3), when one considers the
scalar form factor beyond one loop, see Fig.~\ref{fig:scalarFF}:
it involves isoscalar $s$-wave $\pi\pi$ rescattering, which we found earlier to be strong 
(see Sect.~\ref{sec:a00size}).
A dispersive analysis yields $\langle r^2 \rangle_\sigma \approx 2 \langle r^2\rangle_{EM}$
and a  large curvature term, with the result $\Delta\sigma \approx 15 ~{\rm MeV}$~\cite{GLSsigma}.
The $\sigma$-term is then much smaller,\vspace{-1mm}
\beq
\sigma \approx 45~{\rm MeV} ~, \vspace{-1mm}
\eeq
we now find $y \approx 0.2$ and a 
sizeable, but not outrageously large strangeness contribution to the nucleon mass,
\beq \langle N | m_s \bar ss| N \rangle \simeq 130~{\rm MeV} ~. \eeq


\subsection{\boldmath{$\pi N$} scattering lengths}

In order to improve the knowledge of the $\sigma$-term, it would be useful
to have precise information on $\pi N$ scattering in the threshold region 
(which is closest to the Cheng--Dashen point), or, more precisely, on the scattering lengths,
defined by\vspace{-1mm}
\beq 
a^\pm = \frac{1}{4\pi(1+M_\pi/m_N)} T^\pm \left(s=(m_N+M_\pi)^2\right) ~.
\eeq
In ChPT at leading orders, these have the following expansion~\cite{BKMpiN,nadia}:\vspace{-1mm}
\bea
a^- &=& \frac{{M_\pi}}{8\pi(1+M_\pi/m_N)F_\pi^2} + \Order\bigl({M_\pi^3}\bigr) ~,\no\\
a^+ &=& {0} + \frac{{\Mpi}(-g_A^2+8m_N(-2c_1+c_2+c_3))}{16\pi m_N(1+M_\pi/m_N)F_\pi^2}
+ \Order\bigl({M_\pi^3}\bigr) ~.\vspace{-1mm}
\eea
Numerically we have $a^- = 8.0 \times 10^{-2} M_\pi^{-1}$ plus small higher-order corrections.
In contrast, 
$a^+$ vanishes at leading order $p^1$, receives contributions of several LECs
at $\Order(p^2)$, and can be seen to converge rather badly.
This means that precisely the isoscalar scattering length that would be helpful
in constraining the $\sigma$-term is barely known from ChPT.

Precise experimental information on the scattering lengths can be obtained
from pionic hydrogen and pionic deuterium measurements,
in analogy to the discussion of pionium lifetime for the $\pi\pi$
scattering lengths.
The best values deduced thereof~\cite{MRR06} (see also experimental references therein)
are given by\vspace{-1mm}
\beq
a^- = (8.52 \pm 0.18) \times 10^{-2} M_\pi^{-1} ~,\quad
a^+ = (0.15 \pm 0.22)\times 10^{-2} M_\pi^{-1} ~. \vspace{-1mm}
\eeq
$a^+$ is seen to be very small, and quite sensitive to isospin breaking corrections.


\section{Outlook}

Instead of a summary,
we rather give an outlook on various aspects of ChPT that, 
due to lack of time and space, could not be covered in these lectures.

In the Goldstone boson sector,
of course there are many more processes of high physical interest, e.g.\
various meson decays, further scattering problems etc.  
The whole sector of odd intrinsic parity (``chiral anomaly'', Wess--Zumino--Witten term~\cite{WZW})
has not been touched upon, although it is responsible for such fundamental decays
as $\pi^0 \to \gamma \gamma$ or $\eta \to \gamma \gamma$.
A whole new set of Lagrangian terms  is furthermore needed for 
weak matrix elements, e.g.\ for non-leptonic kaon decays.
We have only in passing hinted at the links to dispersion theory,
and relations of the chiral phenomenology to 
large-$N_c$ arguments have been completely ignored.

In the single-baryon sector, some of the most noteworthy omissions include
the highly interesting topic of isospin violation in $\pi N$ scattering;
the analysis of pion photo-/electroproduction $\gamma^{(*)} N \to \pi N$,
which has proven to be one of the key successes in the development
of baryon ChPT (see \cite{BKMreview} and references therein);
Compton scattering $\gamma N \to \gamma N$ and nucleon polarisabilities.
An important extension of ChPT with baryons is the inclusion of
explicit spin-3/2 ($\Delta$) degrees of freedom (``small-scale expansion''~\cite{SSE}).

The topic of two-and-more-nucleon systems has been covered
in the lectures by Chen, Hanhart~\cite{ChristophProc}, and Machleidt at this workshop.  
The connections to lattice QCD, which are another major point of research at present, 
concerning 
chiral extrapolations, ChPT in a finite volume and at non-zero lattice spacing,
(partially) quenched ChPT have been touched upon in the lectures by Chen.


\subsection*{Acknowledgements}

I would like to thank the organisers of ``Physics and Astrophysics of Hadrons
and Hadronic Matter''
for a wonderful workshop, and in particular
for their overwhelming hospitality during the week in Shantiniketan.
I am grateful to J\"urg Gasser and Ulf-G.~Mei{\ss}ner for useful comments on the manuscript.
This work was supported in parts by the DFG (SFB/TR 16)
and the EU I3HP Project (RII3-CT-2004-506078).


\end{document}